\documentclass{elsart}
\usepackage{natbib,graphicx,fancybox}

\bibpunct{(}{)}{;}{a}{}{,}
\def\bib[#1]{\bibitem[#1]}
\def\cite{\citep}
\def\citett[#1]{\citealt{#1}}
\def\citeextra[#1;#2]{\citet{#2}}

\begin{document}
\begin{frontmatter}
\title{X-ray Spectroscopy of Cooling Clusters}
\author[ad1]{J. R. Peterson\thanksref{corr1}}
\author[ad2]{\& A. C. Fabian\thanksref{corr2}}
\thanks[corr1]{JRP E-mail: jrpeters@slac.stanford.edu}
\thanks[corr2]{ACF E-mail: acf@ast.cam.ac.uk}
\address[ad1]{Kavli Institute for Particle Astrophysics and Cosmology (KIPAC),
  Stanford University, PO~Box 20450, Stanford, CA 94309, USA}
\address[ad2]{Institute of Astronomy (IoA), Cambridge University, Madingley Road, Cambridge CB3~0HA, UK}

\begin{abstract}
We review the X-ray spectra of the cores of clusters of galaxies.  Recent
high resolution X-ray spectroscopic observations have demonstrated a severe
deficit of emission at the lowest X-ray temperatures as compared to that
expected from simple radiative cooling models.  The same
observations have provided compelling evidence that the gas in the cores is
cooling below half the maximum temperature.  We review these results,
discuss physical models of cooling clusters, and describe the X-ray
instrumentation and analysis techniques used to make these observations.  We
discuss several viable mechanisms designed to cancel or distort the expected process of X-ray cluster cooling.
\end{abstract}
\end{frontmatter}

\newpage

\tableofcontents

\newpage

\section{Introduction}

Observations show that the X-ray emission from many clusters of
galaxies is sharply peaked around the central brightest galaxy. The
inferred radiative cooling time of the gas in that peak, where the
temperature drops to the center, is much shorter than the age of the
cluster, suggesting the existence of a cooling flow there \cite{fabian94}. X-ray spectroscopy over the past 5 yr shows that the
temperature drop toward the center is limited to about a factor of
three. Just when the gas should be cooling most rapidly it appears not
to be cooling at all. This is sometimes known as the cooling flow
problem. Careful observations show that gently distributed heat is
required over a radius of up to 100~kpc to balance radiative cooling
in these regions.

The issues of cooling and heating of hot gas have broad relevance to
the gaseous part of galaxy formation and evolution. Brightest cluster
galaxies (BCG) are the most massive galaxies known. Calculations of
the clustering behaviour of cold dark matter predict a power-law mass
distribution for large galaxies whereas the stellar mass observed has
an exponential distribution \cite{benson03}. The truncation of the
stellar mass distribution in massive galaxies is likely due to the
process which stops cooling flows. Simple cooling flows are an
ingredient of semi-analytical models for galaxy formation. The cooling
of hot gas to form stars is essential for the growth of massive
galaxies and cannot be studied directly for isolated systems due to
Galactic absorption. The cores of galaxy clusters offer examples which
can be directly observed. However they do not appear to operate in any
simple manner.  The problem appears to be widespread, from the most
massive clusters to the centers of individual elliptical
galaxies. Heating and cooling problems of hot gases are common in
astronomy, with examples ranging from the interstellar medium of our
own Galaxy to the Solar Corona. 

The diffuse hot ionized plasma in clusters is magnetized which means
that MHD processes may be important \cite{schekochihin04}.

Here we briefly review the main X-ray properties and emission
processes of the intracluster medium (ICM) before showing the X-ray
spectra of cool cores. We then discuss the main solutions which have
been proposed for the cooling flow problem.

\newpage

\section{Clusters of Galaxies}

Clusters are the most massive bound and quasi-relaxed objects in the
Universe. They have total masses of $10^{14}$ to above
$10^{15}~{\rm M_{\odot}}$. The total gas fraction is about 16 per cent with about
13 per cent in the hot ICM and the remaining 3 per cent in stars in
the cluster galaxies. The remaining 84 per cent of the mass is in dark
matter. Gas densities in cluster centers range from as much as
$10^{-1}~{\rm cm^{-3}}$ in peaked clusters to $10^{-3}~{\rm cm^{-3}}$ in the
non-peaked ones. This is in stark contrast to the mean cosmic density
of baryons of about $10^{-8}~{\rm cm^{-3}}$.

\begin{figure}[ht]
\begin{center}
\includegraphics[width=0.45\columnwidth]{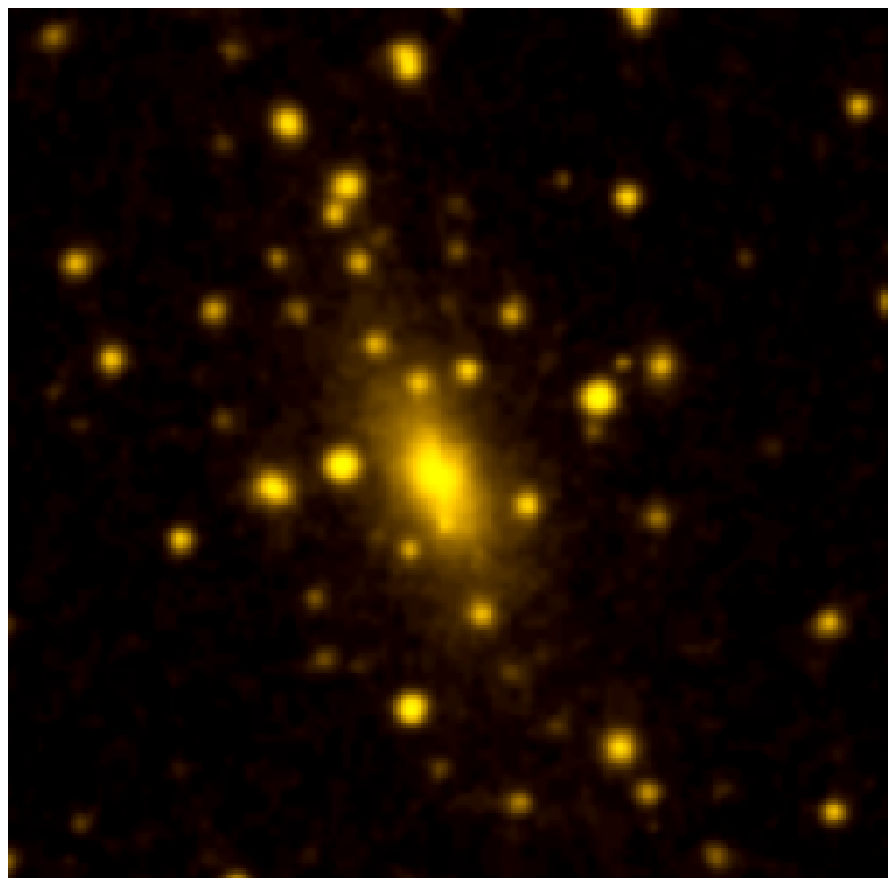}
\includegraphics[width=0.45\columnwidth]{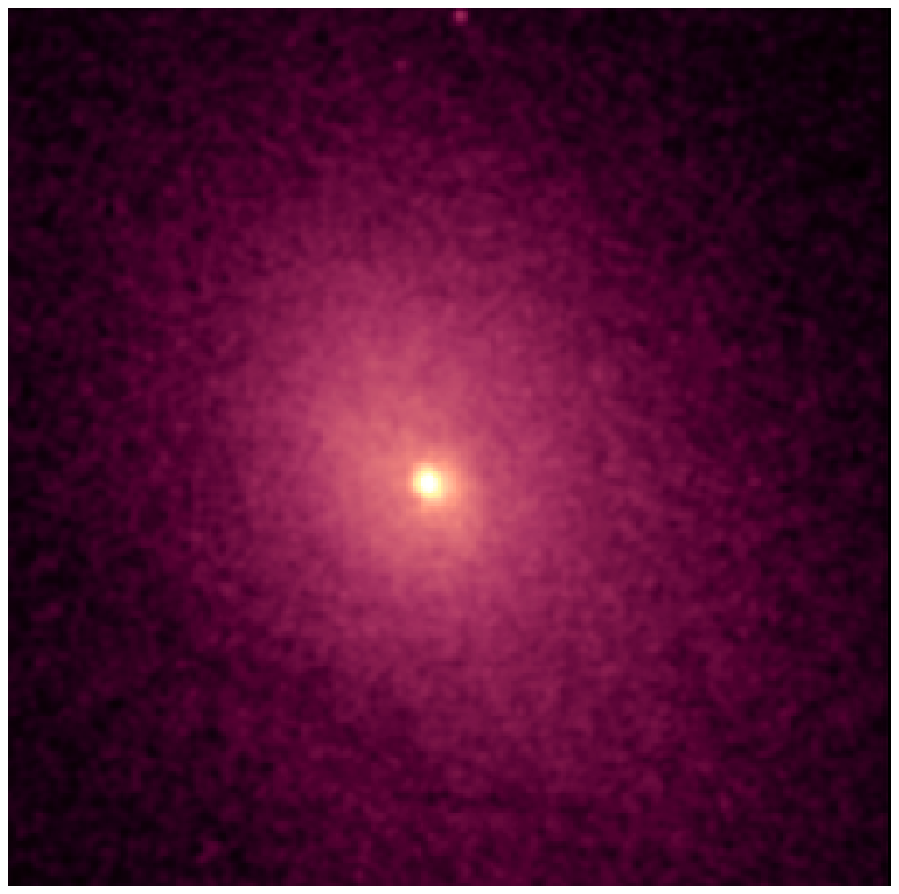}
\end{center}
\caption{Chandra X-ray (left) and DSS optical (right) image of the relaxed
massive galaxy cluster, Abell 2029.  Both images are 4 arcminutes on a side.  Abell 2029 is an extremely regular and putative cooling flow cluster.  The X-ray image
demonstrates how the intracluster medium pervades the space between the
galaxies shown in the optical image.  Figure adapted from http://www.chandra.harvard.edu/ (X-ray: NASA/CXC/UCI/A.Lewis et al. Optical: Pal.Obs. DSS).}
\end{figure}

The characteristic or virial radius, $R_{\rm v}$, of a cluster, defined
from the theory of structure collapse in an expanding universe as
where the mean density of the cluster is 200 times the critical
density of the Universe (i.e. $200\times 3H^2/8\pi G$, with the Hubble
constant at redshift $z$ varying as $H/H_0=\sqrt{\Omega_{\rm m}(1+z)^3
+ 1 -\Omega_{\rm m}})$, is typically between 1 and 3~Mpc.  The gas is
heated by gravitational infall to temperatures close to the virial
temperature $kT\sim GMm_{\rm p}/R_{\rm v}$, which ranges in clusters
from 1--15~keV. The total X-ray luminosities range from about
$10^{43}~{\rm erg~s^{-1}}$ to $10^{46}~{\rm erg~s^{-1}}$. Objects at lower masses and
luminosities are groups which have from a few to tens of member
galaxies as compared with the hundreds of galaxies in a typical
cluster.

Structure formation in the Universe proceeds in a hierarchical manner
with the most massive objects, clusters, forming last, which means
now. They  continue to evolve by the infall of subclusters. The
time since the last major merger is typically about 5~Gyr. About 20
per cent of clusters have had a more recent merger or are undergoing
one. These are not the subject of this review.

Analytic and numerical simulations of cluster formation indicate that
the total X-ray luminosity $L_{\rm X}\propto T^2$ in the absence of gas
cooling and heating. This follows since
the X-ray luminosity is dominated by thermal bremsstrahlung so $L_{\rm
X}\propto n^2 T^{1/2} R_{\rm v}^3$, the mean gas density $n\propto
\frac{M}{R_{\rm v}^3}$ is constant and $T=\frac{M}{R_{\rm v}}$. The
temperature drops monotonically outward (by a factor of up to about
2). Observations instead show $L_{\rm X}\propto T^3$ over the
temperature range 2--8~keV with a wide dispersion at lower
temperatures and a possible flattening above. The simplest explanation
for this result is that the gas has had additional heating of 2--3~keV
per particle \cite{wu00,voit03}. The effect of
such heating is not to increase the temperature by that amount but
mostly to expand the gas (reducing its density and thus X-ray
luminosity).  Such energy is plausibly due to energy output from
active galaxies, i.e. accreting black holes in cluster galaxies.
Alternatively, radiative cooling by removing the low-entropy gas in star
formation may reproduce the relation as well \cite{voit01}.

The gas has generally been enriched to 0.3 of the Solar value by early
supernovae. In relaxed clusters the potential and gas peak on the
BCG. The metallicity often rises to solar or even higher around the
BCG, probably due to SN Ia.

In relaxed, X-ray peaked, clusters the temperature profile is often
inverted in the inner core (i.e. $R<100~{\rm kpc}$) dropping inward as
$T\propto r^\alpha$ with $\alpha\sim 0.3-0.5$. The gas density there
rises as $n\propto r^{-1}$.

The overall profiles of the gas density and temperature depend on the
entropy of the gas and thus on its heating and cooling history, subject
to the equation of hydrostatic equilibrium,
\begin{equation}
{{dp}\over{dr}}=-\rho g~~{\rm or}~~{{d(nkT)}\over{dr}}=-n \mu m_{\rm p}
{{GM(<r)}\over r^2}.
\end{equation} 

where $p$ is the pressure, $\rho$ is the mass density, $n$ is the number
density, $k$ is Boltzmann's constant, $T$ is the temperature, $G$ is Newton's
constant, $g$ is the gravitational acceleration, $M(<r)$ is the enclosed mass
within a radius $r$, $m_P$ is the proton mass, and $\mu m_p$ is the mean mass per particle.  This
equation is used to estimate the total mass profile of  clusters. Massive ones
can act as gravitational lenses for background galaxies as seen in the optical band which provides another means to
measure mass profiles. Agreement between profiles determined by both
methods \cite{allen01c} show that hydrostatic equilibrium
holds well in the main body of relaxed massive lensing clusters and
that any non-thermal pressure there is not dominant.

\newpage
\section{Physics of the Intracluster Medium}

The intracluster medium (ICM) is plasma that is nearly fully ionized
due to the high temperatures created by the deep dark matter
gravitational potential.  Hydrogen and helium, for example, are fully
stripped of their electrons.  Heavier elements have retained only a
few of their electrons in this hot medium. In addition to free
electrons and ions in the plasma, electromagnetic radiation, which is
emitted mostly as X-rays, is created by quantum mechanical
interactions in the plasma.  

The physics of the ICM can be studied as two physical phenomena:
1) the ionized plasma, and 2) the radiation emission processes.  The ionized
plasma is well-described by magneto-hydrodynamic theory on large enough
scales.  The radiation emission processes are governed by equations for X-ray
emission from a collisionally-ionized plasma.  We can treat
these phenomena separately because the intracluster medium is mostly
optically-thin, (i.e. the radiation almost completely escapes without
interacting with the plasma).  Later in the next section, we will show that
this is the case.

In \S3.1, we describe how X-rays are produced in the intracluster
medium.  In section \S3.2, we motivate magneto-hydrodynamic theory as the
description of the plasma.  In \S3.3, we unify the two theoretical
descriptions to produce what we refer to as the ``standard cooling-flow
model''.  This model will serve as the basis for the physical description of
the cores of clusters of galaxies.  In later sections, we describe the
puzzling observations that seem to agree with many expectations of this model,
but disagree strongly with other aspects of the model.

\subsection{X-ray Emission from Collisional Plasmas}

Ionized plasmas produce copious amounts of X-rays.  The emission of
X-rays has two important consequences.  First, it allows us to observe
the intracluster medium by detecting those X-rays.  Furthermore, since
most of those X-rays do not interact between their emission and their
detection in X-ray telescopes it allows us to study the intracluster
medium in an unperturbed state.  Through X-ray spectroscopy and
imaging, we can measure several physical quantities, such as the
temperature and density, at various positions in the cluster.

The second important consequence is that the emission of these X-rays
will tend to cool the plasma.  A significant quantity of energy is
carried away by the X-rays as they escape the cluster.  The emission
of these X-rays was thought to set up a non-linear process of
excessive cooling in cluster cores that is broadly termed a ``cooling
flow'', a major subject of this work.  In the following, we describe
the process of X-ray emission from collisionally-ionized plasmas and
show how it relates to the intracluster medium.

\subsubsection{Coronal Approximation}

The emission of X-rays from ionized atoms in a plasma can be quite complex.
Fortunately, we can make use of several approximations to simplify the
emission processes.  The emission of X-rays in the ICM is well-described by the
coronal approximation \cite{elwert52,mewe93}.   These approximations describe
optically-thin plasmas in collisional equilibrium.  Collisional equilibrium
occurs when electron collisional ionization processes are balanced precisely
by recombination processes.  The coronal approximation, as the name implies,
was originally developed for study of the Solar Corona, but the condition
also applies to gas in clusters of galaxies, as well as hot gas in elliptical,
starburst galaxies, and older supernovae remnants.  In this approximation,
there are three important conditions that specify the thermodynamic state of
the free electrons, ions, and photons in the plasma, as well as the electron distribution within each ion.

The first approximation is that the photons are assumed to be free and do not
interact with either the electrons or the ions after they are created.  This
has the importance consequence that photo-ionization processes (ionizing atoms
by photons) and photo-excitation processes (raising an electron in an atom to
an excited level) are far less frequent that electron collisional ionization
and excitation processes.  The radiation densities are low enough in clusters
of galaxies for this condition to be met, except possibly for resonant
scattering in some strong emission lines with high oscillator strength \cite{gilfanov87}.

The second approximation is that the atoms can be treated as if their
electrons are all in their ground state rather than having a Boltzmann
distribution as is common in LTE (local thermodynamic equilibrium) gases.
This is true if there is a low enough electron and radiation density, so that
the excitations that are density dependent are less frequent than the
radiative decays, which are quite fast for X-ray energies.  The radiation
density is low as discussed above.  This condition is also met for electron
densities below $10^{10}~\mbox{cm}^{-3}$ for even slowly decaying metastable
states.  Densities in clusters are at most $10^{-1}~\mbox{cm}^{-3}$.  

The final approximation is that the plasma is locally relaxed to a Maxwellian
distribution around a common electron temperature, T.  The free electrons and
ions are assumed to have obtained a common temperature.  This is only valid if
typical dynamical time-scales, such as the time it takes the plasma to cool,
is much longer than the time scale for sharing energy between electrons and
ions, such as the electron recombination time scale or time scale between
Coulomb collisions.   If this is true, then collional equilibrium is achieved
in which ionizations are balanced by recombinations.  This assumption is valid
in the cores of cluster, but may break down in the outer regions of clusters
where the density is lower and the plasma may still be in the process of
ionizing.  One might also worry that this assumption may break down in the
cores of clusters if there is a complex multi-phase distribution of plasma
temperatures due to thermally unstable pockets of plasma each with their own
temperature.  Generally if such a situations exists, however, it is likely
that each pocket will locally achieve a common temperature.  An electron
penetrating a cloud of a different temperature will interact and achieve a
Maxwellian distribution.  It is only with complex and rapidly mixed interfaces
that the time scale of injection of a new cloud is much shorter than
$\frac{\lambda_e}{v_e}$, where $v_e$ is the electron velocity, and $\lambda_e$
is the electron mean free path, that the collisional equilibrium would be violated.

\subsubsection{Ionization Balance}

The fraction of atoms at a given charge state is determined by a balance of
ionizations and recombinations.  The exact balance is fixed by a coupled set
of differential equations that relate the ionization and recombination
processes between neighboring charge states.  The equations are of the form,

\begin{equation}
\frac{dn_i}{dt} = -I_i n_e n_i - R_i n_e n_i + I_{i-1} n_e n_{i-1} +  R_{i+1} n_e n_{i+1}.
\end{equation}

\noindent
where $n_e$ is the electron number density, $n_i$ is the density of atoms in
the $i$th charge state, $I_i$ is the ionization rate out of the $i$th charge
state, and $R_i$ is the recombination rate out of the $i$th state.  We have
ignored spatial diffusion of the ionization balance.  Note, that
it is customary to express the density of a given ion relative to the number
density of hydrogen, $n_H$.  We then define $a_i$ as the relative abundance
and the fraction of atoms in a given charge state as $f_i$.  Then we have,
$n_i= n_H f_i a_i$, where $n_H a_i$ drops out of the above equations.
Collisional equilibrium assumes that a steady-state has been achieved, so that
the left hand side is set to zero.  The equations therefore simplify to
equations of the form,

\begin{equation}
0 = -I_i f_i + R_{i+1} f_{i+1}
\end{equation}

\begin{figure}[ht]
\begin{center}
\includegraphics[width=0.9\columnwidth]{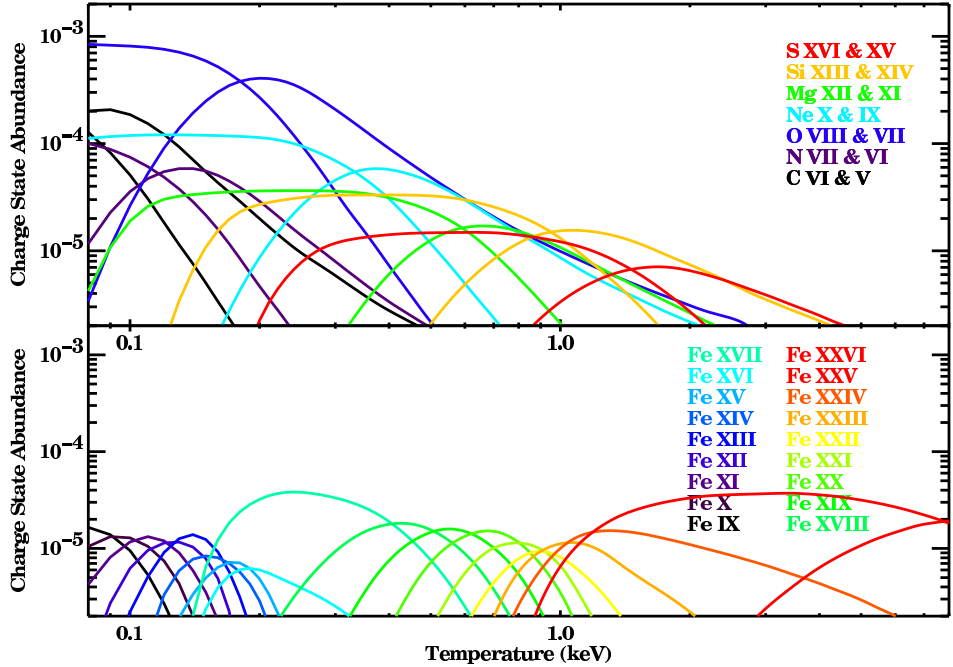}
\end{center}
\caption{\label{fig:ionabund} The charge state abundance (elemental abundance times fraction ionic
abundance) of various ions as a function of temperature.  The top
panel shows helium-like and hydrogen-like charge states of various low Z
atoms.  The bottom panel shows iron ions having the outer electron in the K,
L, and M shell.  The bottom panel indicates how the measurement of various
ions in the iron series is a sensitive probe of whether plasma at a given
temperature exists.  Figure uses data from \citett[arnaud92].}
\end{figure}

\noindent
  Collisional equilibrium will eventually be achieved if a plasma
remains undisturbed for a long enough period of time (i.e., the inverse of the
recombination rate).  Several calculations have been done to determine this
ionization balance as a function of the electron temperature
\cite{jordan69,arnaud85,arnaud92,mazzotta98}.  One such calculation is
shown in Figure~\ref{fig:ionabund}.

Once the ionization balance is determined the X-ray spectrum can be calculated
by considering the various radiation processes.  The most important processes
are bremsstrahlung and the K and L shell transitions for the
discrete line emission.

\subsubsection{Bremsstrahlung and other Continuum Processes}

When Hydrogen is ionized above temperatures of $2\times 10^4$ K,
copious amounts of bremsstrahlung emission are produced.
Bremsstrahlung radiation results from the accelerations of the free
electrons in the Coulomb field of an ion.  The spectrum is roughly
independent of energy below the energy equal to $kT_e$, where $T_e$ is
the electron temperature and $k$ is Boltzmann's constant.  A rough
approximation of the power per energy per volume radiated by
bremsstrahlung is given by the equation below, \begin{equation}
\frac{d^2P}{dVdE} \approx
10^{-11}~n_e~n_H~T^{-\frac{1}{2}}e^{-\frac{E}{kT}}~\mbox{cm}^{-3}~\mbox{s}^{-1}
\end{equation} \noindent where $n_e$ is the electron density, $n_H$ is
the hydrogen density, $E$ is the photon energy, $k$ is Boltzmann's
constant, and $T$ is the electron temperature in Kelvin. The total
power radiated is therefore, \begin{equation} \frac{dP}{dV} \approx
10^{-27}~n_e~n_H~T^{\frac{1}{2}}~\mbox{ergs}~\mbox{cm}^{-3}~\mbox{s}^{-1}
\end{equation} 
\noindent In addition to bremsstrahlung, bound-free
emission (the capture of a free electron to a bound state) and
two-photon emission (which occurs most frequently following a
collisional excitation of Hydrogen to the 2s level) are also
significant source of continuum radiation.  They both modify the shape
of the continuum emission.

\subsubsection{Discrete Line Emission}

Discrete line emission is formed by a number of atomic processes.  Such line
emission is the most important tool for the X-ray spectroscopist.  The most
important atomic processes are collisional excitation, radiative
recombination, dielectronic recombination, and resonant excitation.
Generally, the processes have been incorporated in a number of publically
available and well-tested codes that are used to study collisionally-ionized
spectra.  

The strength of an emission line is determined by the excitation and
recombination rates, which are proportional to the integral of the velocity
times the cross-section for a particular process over a Maxwellian
distribution.  The volume emissivity for a given emission line is calculated
by equations of the form,

\begin{equation}
\epsilon = n_e n_i \left( C^{II} f_{i-1} + C^{E} f_i + \alpha^{R} f_{i+1} \right)
\end{equation}
\noindent
where $C^{II}$ is the rate of inner shell ionization processes, $C^{E}$ is the
sum of collisional excitation processes, and $\alpha^R$ is the sum of
recombination processes.  Several well tested codes have been developed to
calculate the emergent spectrum by assuming an ionization calculation and
including a set of excitation and recombinations rates.  These include the
Raymond-Smith \cite{raymond77}, MEKA \cite{mewe85,mewe86,kaastra92,mewe95},
MEKAL \cite{liedahl95}, and APEC codes \cite{smith01}.  These codes are
compilations of the results of more detailed atomic codes, which solve the
Dirac equation either by the distorted wave approximation \cite{barshalom01,gu03}
or R-matrix methods \cite{berrington95}.  The number of transitions and the
accuracy of the detailed processes limits the results.  Extensive laboratory
work has been applied to verify wavelengths \cite{brown98,brown02} and
cross-sections \cite{gu99,chen05} of the transitions.

A number of common spectral transitions occur in most X-ray spectra.
In fact, despite the complexity implied by the above discussion there
are usually only a couple of dozen strong transitions that are used to
determine most of the information that can be extracted from the X-ray
spectrum.  Several of the important emission line blends are shown in
Table 1.

\begin{table}
\caption{Important line blends in Cluster X-ray Spectra}
\[
\begin{array}{p{0.20\columnwidth}llll}
\hline
\noalign{\smallskip}
\rm{Ion$^{a}$} & \rm{Wavelengths} & \rm{Energies} & \rm{Temperature}  \\
\hline
\noalign{\smallskip}
  & \mbox{\AA} & \rm{keV} & \rm{keV^{b}} \\
\hline
\noalign{\smallskip}
  Fe~XXVI & 1.8 & 6.97 & > 3.0   &  \\
 Fe~XXV & 1.9,1.9 & 6.70, 6.63 &1.0 \rightarrow 8.0  \\
Fe~XXIV  & 10.6, 11.2 & 1.17, 1.11 & 0.9 \rightarrow 4.0 \\
Fe~XXIII & 11.0, 11.4, 12.2 & 1.13,1.09,1.02 & 0.8 \rightarrow 2.0 \\
Fe~XXII  & 11.8, 12.2 & 1.05,1.02 &  0.6 \rightarrow 1.5 \\  
Fe~XXI	 & 12.2, 12.8 & 1.02,0.97 & 0.5 \rightarrow 1.0  \\
Fe~XX    & 12.8, 13.5 & 0.97,0.92 & 0.4 \rightarrow 1.0  \\
Fe~XIX   & 13.5, 12.8 & 0.92,0.97 & 0.3 \rightarrow 0.9  \\ 
Fe~XVIII & 14.2, 16.0 & 0.87, 0.77 & 0.3 \rightarrow 0.8 \\
Fe~XVII & 15.0, 17.1 & 0.83,0.73 & 0.2 \rightarrow
 0.6 \\ 
        & 15.3, 16.8 & 0.81,0.73 & \\ 
S~XXVI & 4.7 & 2.62 &  > 1.0 \\
S~XXV & 5.1, 5.0 & 2.43, 2.46 & 0.3 \rightarrow 1.0 \\
Si XIV & 6.2 & 2.00 & >1.0 \\
Si XIII & 6.6, 6.7 & 1.87, 1.84 &  0.2 \rightarrow 1.0  \\
Mg XII & 8.4  & 1.47 & >0.7 \\
 Mg XI  & 9.2, 9.3 & 1.35,1.33 & 0.1 \rightarrow 0.6\\
 Ne X & 12.2  & 1.02 & >0.4  \\
Ne IX & 13.5, 13.7 & 0.92,0.90 & 0.1 \rightarrow 0.3\\
 O VIII & 19.0, 16.0 & 0.64 & >0.2 \\
 O VII & 21.6, 22.0  & 0.57, 0.56 & 0.1 \rightarrow 0.2  \\
N VII & 24.8 & 0.50& >0.1  \\
C VI & 33.7 & 0.37 & >0.1  \\
\noalign{\smallskip}
\hline
\end{array}
\]
\begin{list}{}{}
\item[$^{a}$] Note this line transition list is somewhat crude since we have
 chosen to tabulate line blends rather than actual transitions, but matches
 well with the quality of the observations.
\item[$^{b}$] The temperature range is calculated to roughly show the temperature range where the emissivity of a given ion blend is within an order of magnitude of its peak emissivity.
\end{list}
\end{table}

\subsubsection{X-ray Cooling Function and Emission Measure Distribution}

The X-ray cooling function is calculated by integrating the emission from all
processes and weighting by the energy of the photons.

\begin{equation}
\Lambda \left( T, Z_i \right) = \int_0^{\infty} dE~E~\frac{d \alpha}{dE}(E,
  T,Z_i)
\end{equation}

\noindent
where $\frac{d\alpha}{dE}$ is the energy dependent line power (or continuum
power).  The cooling function relates the total amount of energy emitted per
volume for a given amount of plasma with a given temperature and emissivity.
The cooling function has been compiled in various tables \cite{boehringer92,sutherland93}.  

\begin{figure}[ht]
\begin{center}
\includegraphics[width=0.9\columnwidth]{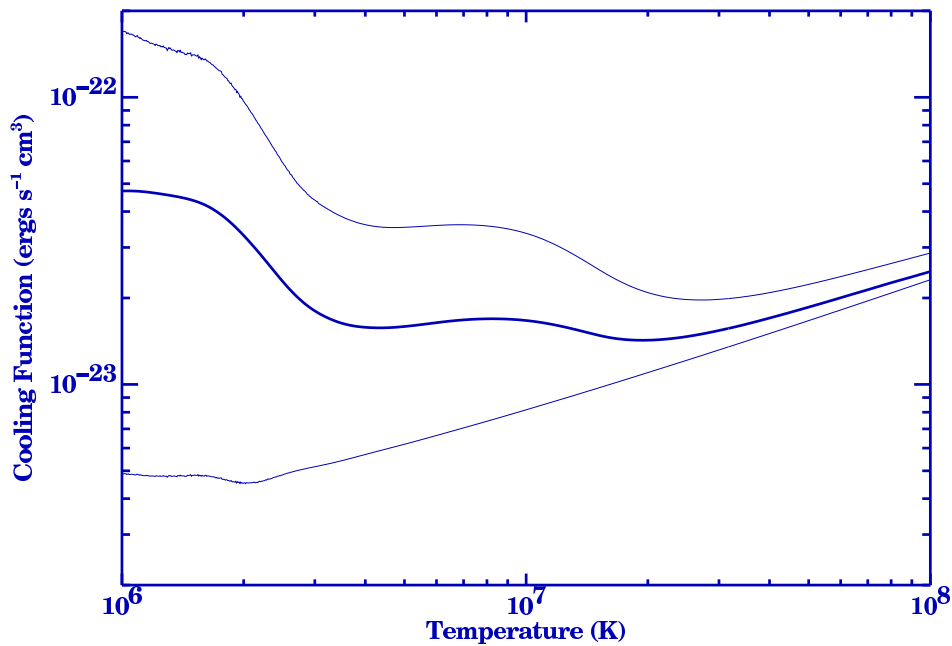}
\end{center}
\caption{The radiative cooling function is shown for solar abundances (top curve), one-third solar abundances
(middle), and pure Hydrogen and Helium (bottom).  The X-ray region from
$10^6$ to $10^8$ K is dominated by bremsstrahlung at high temperatures as well
as significant contribution from line emission at lower temperatures.  Below
$10^6$ K in the UV temperature range, the cooling function rises
significantly.}
\end{figure}

The relative distribution of plasma at a set of temperatures is often
expressed in terms of an emission measure distribution.  The differential
emission measure, $\frac{dEM}{dT}$ is defined by

\begin{equation}
\frac{d\epsilon}{dE} = \int_0^{\infty} \frac{dEM}{dT}~\frac{d\alpha}{dE}(E,T)~dT~.
\end{equation}

\noindent
where $\frac{d\epsilon}{dE}$ is the energy-dependent emissivity and the
integral is over all temperatures.  It is also convenient to express the distribution of plasma temperatures in
terms of the differential luminosity, which is defined by 

\begin{equation}
\frac{dL}{dT} = \frac{dEM}{dT} \Lambda(T)~.
\end{equation}

\subsubsection{Cooling Time}

The cooling time of an optically-thin plasma is the gas enthalpy divided by the
energy lost per unit volume of the plasma.  The gas enthalpy is $\frac{5}{2} n
k T$ and the energy lost per volume is the electron density squared times the
cooling function.  The cooling time can then be written as,

\begin{equation} t_{cool} \equiv \frac{\frac{5}{2} n k T}{n^2 \Lambda}
\approx t_H T_8 \Lambda_{-23}^{-1} n_{-2}^{-1} \end{equation}
\noindent where $t_H$ is the age of the universe (13.7 Gyr), $T_8$ is
the temperature in units of $10^8$ K, $\Lambda_{-23}$ is the cooling
function in units of $10^{-23}~\mbox{ergs}~\mbox{cm}^{3}~s^{-1}$, and
$n_{-2}$ is the density in units of $10^{-2}$ particles
$\mbox{cm}^{-3}$. We used the gas enthalpy per volume, $\frac{5}{2} n
k T$ instead of the thermal energy per volume $\frac{3}{2} n k T$
since the plasma is compressed as it cools which therefore effectively
raises its heat capacity by a factor of $\frac{5}{3}$.  Therefore,
X-ray plasma with gas density above $10^{-2}~\mbox{cm}^{-3}$ has had
sufficient time to cool.  In the cores of cooling clusters the cooling
time approaches cooling times below $5 \times 10^8~\mbox{yr}$.  If the
gas was undisturbed, it would have a chance to cool several times.
Note that as gas cools at constant pressure (due to the weight of
overlying gas) then the rise in density as the temperature drops means
that $t_{\rm cool}$ becomes shorter and shorter.

\subsubsection{Optical Depth, Resonance Scattering, and Opacity}

The optical depth for photons of a given wavelength, $\tau$, is the product of
the column density of a particular ion, $N_i$ and the cross-section of a
particular process.  The cross-section is a function of energy.  The
column density is the line integral of the ion density, $n_i$, which is a
function of the spatial position along the line of sight.

The intracluster medium is, for the most part, optically-thin to its own
radiation (i.e. photons escape once they are emitted without interacting
again).  Photons with energies close to certain resonance transitions,
however, can scatter several times before leaving a cluster.  The optical
depth to a resonance transition is given by

\begin{equation}
\tau=\int dl n_i f_i \frac{\pi e^2}{m_e  c} \frac{exp^{-\frac{(\lambda-\lambda_i)^2}{2
      \sigma_i^2}}}{\sqrt{2 \pi \sigma_i^2}}
\end{equation}

\noindent
where $n_i$ is the ion density, $f_i$ is the oscillator strength of the
transition, $m_e$ is the electron mass, $\lambda_i$ is the wavelength of the
transition, $c$ is the speed of light, $e$ is the electron charge, and
$\sigma_i$ is the line width given by,

\begin{equation}
\sigma_i= \lambda_i \sqrt{\frac{kT}{\left(\frac{A}{1+\frac{5}{3}M^2\frac{A}{\mu}} \right) m_p c^2}}
\end{equation}

\noindent where $A$ is the atomic mass number, $\mu$ is the mean mass
per particle, $M$ is the Mach number of the turbulence or gas motions
in the plasma, $k$ is Boltzmann's constant, and $T$ is the
temperature.  The above expression includes both the thermal
broadening of the line as well as turbulent broadening.

If a significant quantity of lowly ionized matter exists along the
line of sight, X-rays can be absorbed and re-emitted at lowly ionized
longer wavelengths.  This situation occurs frequently for absorption
from neutral gas in the Milky Way Galaxy, but could also occur from
gas trapped in the cluster potential.  This subject is discussed in
detail later, but we note that understanding the role of absorption
often has a significant effect on the interpretation of the soft X-ray
spectrum. This is particularly true at low resolution.  Helium K shell
absorption at low energies and Oxygen K shell absorption at 23.5
$\mbox{\AA}$ are the largest contributors to the opacity from a
neutral absorber and produce absorption edge features in the spectrum.

\subsection{Magneto-hydrodynamics}

The plasma of the intracluster medium can be described by
magneto-hydrodynamics on large scales.  Several assumptions underlie this
statement.  First, the ICM is optically-thin so we do not have to include
radiative forces in the theory.  Second, we assume the ICM is
non-relativistic.  This is true since the sound speed is at most a few thousand
kilometers per second.   Third, we assume the ICM is nearly fully-ionized,
which is true since Hydrogen consitutes most of the plasma.  Finally, the plasma can be
treated as a fluid using continuous fields if the plasma parameter is small so
that collective processes dominate.  The plasma parameter is defined by 

\begin{equation}
g \equiv \frac{1}{n \lambda_D^3}= 8 \times 10^{-3} n^{\frac{1}{2}} T^{-\frac{3}{2}}
\end{equation}

\noindent
where $\lambda_{D}$ is the Debye length, $n$ is the density in units of
$\mbox{cm}^{-3}$, and $T$ is the temperature in K.  The ICM plasma parameter
is in fact extremely small.

Magnetohydrodynamics (MHD) in this context is described by the following variables:  the fluid velocity, ${\bf v}$, the temperature, $T$, the density,
$\rho$, and the magnetic field, ${\bf B}$, and the
dark matter density, $\rho_{DM}$.  These quantities are related to one another
by the full set of MHD equations, which assume local thermodynamic
equilibrium of these quantities.  The transport properties of the fluid can be
written in terms of the viscosity, $\eta$,
conductivity, $K$, and resistivity, $\lambda$.  

We also adopt the following notation.  T is the energy per particle; whereas $T_K$ is the temperature in
Kelvin.  They are related to one another by the
relation, $T= {k T_K}/{\left(\mu m_p \right)}$.  $k$ is Boltzmann's
constant, the mass of the proton is $m_p$, and the mean mass per particle is
$\mu m_p$.  Additionally, the cooling luminosity is usually defined in terms of the
electron, $n_e$ and hydrogen, $n_H$, number densities, which are related to the
fluid mass density by, $n_e = {n_H}/{1.19} = {\rho}/{\left(\mu m_p\right)}$.

Below are the 5 MHD equations.  

\noindent
\fbox{\parbox{0.98\columnwidth}{ \small
\begin{equation}
\frac{\partial {\rho}}{\partial {t}}  +\nabla \cdot \left( \rho {\bf
v} \right)  =  0
\end{equation}
\begin{eqnarray}
\rho \frac{\partial v_i}{\partial t}  + \rho \left( {\bf v} \cdot \nabla \right) {v_i}   =  
\rho \frac{\partial \Phi}{\partial x_i} - \frac{\partial}{\partial x_i} \left( \rho T + \frac{B^2}{8 \pi} \right) + \frac{ \left( {\bf B} \cdot \nabla \right) { B_i}}{4 \pi} \nonumber \\
   + \frac{\partial}{\partial x_j} \left[ \eta \left( \frac{\partial
    v_i}{\partial x_j} + \frac{\partial v_j}{\partial x_i} - \frac{2}{3}
  \delta_{ij} \nabla \cdot {\bf v} \right) \right] 
\end{eqnarray}
\begin{equation}
\frac{\partial {\bf B}}{\partial t}  =  \nabla \times \left( {\bf v} \times {\bf
    B} \right) - \nabla \times \left( \lambda \nabla \times {\bf B} \right)
\end{equation}
\begin{eqnarray}
\frac{3}{2} \rho \frac{\partial T}{\partial t}   + \frac{3}{2} \rho {\bf
v} \cdot \nabla T + \rho T \nabla \cdot {\bf v}  =  \nabla \cdot \left( K \nabla T_K \right) + \frac{\lambda}{4 \pi} \left( \nabla \times {\bf B} \right) \cdot \left( \nabla \times {\bf B} \right) \nonumber \\
    +2 \eta \left( \frac{\partial v_i}{\partial x_j} + \frac{\partial
v_j}{\partial x_i} \right)^2 - \frac{2}{3} \eta \left( \nabla \cdot {\bf
v} \right)^2 - n_e n_H \Lambda (T,Z) + \rm{interactions} 
\end{eqnarray}
\begin{equation}
\nabla^2 \Phi   =  4 \pi G(\rho+\rho_{DM})
\end{equation}
}}

The first equation, the mass conservation equation, is a statement that the
total mass of the fluid is constant.  The second equation, the Navier-Stoker equation, enforces
momentum conservation.  This equation relates the momentum of
the fluid (left hand side) to the gravitational compression (first term on the
right hand side), the thermal and
magnetic pressure gradients (second and third term), tangetial magnetic
transport (fourth term), and viscous forces (last terms).

The evolution of
the magnetic field follows the induction equation, the third equation.
The first term on the right hand side generates the magnetic field due to
plasma motion, and the final term dissipates the magnetic flux due to magnetic
reconnection.  

The energy equation, the fourth equation, expresses the balance between
heating and cooling terms.  The left hand side describes the energy content of the plasma (first and
second term) as well as its compression (third term).  The right hand side
contains a conduction term (first term), a magnetic dissipation term (second
term), viscous heating terms (third and fourth term), and the energy lost due
to radiative cooling.  In addition, this equation could include interactions
between this plasma and other matter, such as dust, cosmic rays, or dark
matter.  Finally, the gravitational field is set by the fifth equation
where both the dark matter and plasma contribute to the gravitational field.

The transport coefficients for the intracluster medium is the subject of much
theoretical work.  The values of these coefficients for a ionized plasma with
no magnetic field was worked out by \citeextra[Spitzer;spitzer62] using kinetic theory.
These values are given by:

\noindent
\fbox{\parbox{0.98\columnwidth}{ \small
\begin{eqnarray}
K       & \approx & 10^{-5}~T^{5/2}~{\rm ergs~cm^{-1}~s^{-1}~deg^{-1}}	\\
\eta    & \approx & 10^{-17}~T^{5/2}~{\rm g~cm^{-1}~s^{-1}}	\\
\lambda & \approx & 7\times 10^{12}~T^{-3/2}~{\rm cm^{2}~s^{-1}}.
\end{eqnarray}
}}

\noindent
A tangled magnetic field due to MHD turbulence, e.g. \citett[goldreich95], could
modify these coefficient considerably and there has been considerable debate
in this subject \cite{tao95,chandran98,narayan01,maron04}.  We will return to this subject in later sections.

\subsection{Cooling flows}

The radiative cooling time in the cores of at least two thirds of low
redshift \cite{peres98} and moderate redshift \cite{bauer02}
clusters is less than 10~Gyr and for one third it is less than about
3~Gyr. The energy loss is directly due to the observed X-ray emission
with no major bolometric correction. If there is no heating to
compensate the cooling then a cooling flow occurs in these regions. In
order to understand what the 'cooling-flow problem' is, why heating
is required, how a 'residual flow' might operate and what happens when
heating is not operating, we now briefly examine cooling flows.

\subsubsection{Single-phase flows}

The radiative cooling time $t_{\rm cool}$ at tens of kpc radius 
in a cluster always
exceeds the gravitational dynamical time so cooling leads to a slow,
subsonic inflow there. The flow causes the density to rise and
so maintain the pressure, which is determined by the weight of the
overlying gas. 

We can simplify the MHD equations significantly for the simple case of
an
unmagnetized single-phase subsonic flow.  We also ignore any terms with
conduction, viscosity, and resistivity.  If we simplify the previous equations
and rewrite the LHS of the energy equation based on the mass equation,

\noindent
\fbox{\parbox{0.98\columnwidth}{ \small
\begin{equation}
\frac{\partial {\rho}}{\partial {t}} + \nabla \cdot \left( \rho {\bf v} \right)  =  0
\end{equation}
\begin{equation}
\rho \frac{\partial v_i}{\partial t}  + \rho \left( {\bf v} \cdot \nabla \right) {v_i} =  
\rho \frac{\partial \Phi}{\partial x_i} - \frac{\partial \left( \rho T \right)}{\partial x_i}  \nonumber \\
\end{equation}
\begin{equation}
\rho  \left[ \left( \frac{\partial}{\partial t} + {\bf
v} \cdot \nabla \right) \frac{5}{2} T - \frac{1}{\rho} \left( \frac{\partial}{\partial t} + {\bf
v} \cdot \nabla \right) p \right] =   - n_e n_H \Lambda (T,Z)
+ \rm{interactions} 
\label{eqn:energy}
\end{equation}
\begin{equation}
\nabla^2 \Phi   =  4 \pi G(\rho+\rho_{DM}).
\end{equation}
}}

We now assume a spherical geometry and assume that the system is in a
steady-state such that the partial time derivatives can be ignored, and assume
that the flow is subsonic, then terms of order $v^2$ can be ignored.  We combine the second equation with the
third, and define $\dot{M}$ in the first equation, and obtain,

\noindent
\fbox{\parbox{0.98\columnwidth}{ \small
\begin{equation}
\dot{M} \equiv 4 \pi r^2 \rho v =  \mbox{constant}
\end{equation}
\begin{equation}
\rho \frac{d \Phi}{dr} = \frac{d \left( \rho T  \right)}{dr}  \nonumber \\
\end{equation}
\begin{equation}
\rho  v \frac{d}{dr} \left( \frac{5}{2} T - \Phi \right)  =   - n_e n_H \Lambda (T,Z) + \rm{interactions} 
\end{equation}
\begin{equation}
\frac{1}{r^2} \frac{d}{dr} \left( r^2 \frac{d \Phi}{dr} \right)  =  4 \pi G(\rho+\rho_{DM}).
\end{equation}
}}

When the gravitational field is important, then the temperature
approaches a critical solution which 'follows the gravitational
potential' \cite{fabian84,nulsen86} as can be seen from the
hydrostatic equation for a power law solution.  Essentially the flow
settles down to a temperature profile close to the local virial one.
In an NFW \cite{navarro97} potential where the inner power-law part
has dark matter mass density varying as $\rho \propto
r^{-1}$ this means $T\propto r$. The
temperature then flattens to $T \sim 1~{\rm keV}$ within the inner $\approx$
10~kpc where the gravitational potential of the central galaxy, assume
isothermal, dominates. The temperature finally collapses at the
center, before which the flow may go supersonic (the inertial velocity
term is then needed in the above momentum equation).  Over most of the
region $r/v\approx t_{\rm cool},$ as seen in the energy equation,
which varies as $T^{\frac{1}{2}}/n$ for bremsstrahlung. Using the mass flow
equation to substitute for $v$, $n\propto
r^{-\frac{5}{4}}$ in the NFW case which leads to a steeper surface
brightness profile (the emissivity is proportional to $n^2$) than
observed.

Even with the King potential commonly used before the 1990s, it was realized
that the central surface brightness was too steep to match the
observed profiles \cite{fabian84}. Interpreted as a cooling flow, the
data indicated that the mass flow rate increased with radius (roughly
as $\dot M \propto r$ out to the cooling radius $r_{\rm c}$ where
$t_{\rm cool}\sim 10^{10}~{\rm yr}$). Such a situation requires that matter
is cooling out over a range of radii, which was explained by gas with
a range of densities and thus cooling times occurring at a given
radius.  Therefore, more generic models than a simplified spherical
single phase flow had to be considered.

\subsubsection{Thermal Instability and Multiphase flows}

The above discussion of single phase flows says little about what happens on
small scales.  In particular, when cooling begins and what size
perturbation leads to the largest growth rate and whether the multiphase flows
could develop like the data seemed to indicate are an open question.  There
has also considerable
effort to understand whether the reservoir of heat in the outer regions of
clusters can be transferred to the center, which could effectively stabilize any
initially thermally unstable parcels of gas.

\citeextra[Field;field65] originally discussed the origin of thermal instability
due to the emission of radiation.  He found that for all X-ray temperatures the gas
is thermally unstable and the growth rate is fastest on the smallest scales.
He further found that the small scale perturbations are damped by conduction
so that the growth rates will be faster on somewhat larger scales.
A number of authors \cite{malagoli87,white87a,balbus88,loewenstein89,balbus89}
studied thermal instability in the context of gravitational field.
\citeextra[Balbus;balbus91] noted that some of the thermal instability arguments are inapplicable in a
cluster gravitational potential.   It is possible that overdense parcels of plasma can come
to equilibrium at a lower adiabat deep in the potential \cite{cowie80}.
\citeextra[Kim \& Narayan;kim03b] argue that the radial modes are unstable
even in the presence of conduction.

\citeextra[Nulsen;nulsen86] and \citeextra[Thomas;thomas87], however,
considered the development of perturbations into cooling flows and
discussed the possible importance of magnetic fields in pinning
parcels of plasma to the general hydrodynamic flow.
\citeextra[Loewenstein;loewenstein90] discussed the importance of the
magnetic field in altering the instability conditions by effectively
eliminating the buoyancy problem by \citeextra[Balbus;balbus89] with
magnetic stresses.  \citeextra[Balbus;balbus91] later confirmed these
instabilities but stressed the importance of inefficient conduction
for these conditions.  However a particular, and not explained,
spectrum of density perturbations is required to obtain the inferred
relation $\dot M\propto r$ \cite{thomas87,tribble89}.
\citeextra[Binney;binney04] argues that multiphase flows do not occur
in real clusters.

\subsubsection{Standard Multiphase Cooling Flow Model}

The standard multiphase model simplifies the physics of the flow by
assuming that radiative cooling dominates the flow and then looking at
the relative amount of radiation emitted at each temperature over the
cooling flow volume. This can be seen as a starting point to testing
the idea that cooling flows are operating in the cores of clusters. If
we restore Equation \ref{eqn:energy} to the full time-derivative form
and then integrate over volume and neglect any heating or
additional cooling we obtain, 

\begin{equation}
\rho  \left( \frac{5}{2} \frac{dT}{dt} - \frac{1}{\rho} \frac{dp}{dt} \right)
dV =   - n_e n_H \Lambda (T,Z) dV.  
\label{eqn:smcf1}
\end{equation}

We define a mass loss rate per time, $\dot{M}$, and a differential X-ray
luminosity, $dL_x$, according to

\begin{equation} 
- \dot{M} \equiv \rho \frac{dV}{dt};~~~~dL_x \equiv n_e n_H \Lambda (T,Z) dV.
\end{equation}

Then the Equation~\ref{eqn:smcf1} simplifies to,

\begin{equation}
\dot{M}  \left( \frac{5}{2} dT - \frac{1}{\rho} dp \right) =  dL_x.
\end{equation}

This can be rewritten as,

\begin{equation}
\frac{dL_x}{dT}= \dot{M} \left( \frac{5}{2}
- \frac{1}{\rho} \frac{dp}{dT} \right).
\label{eqn:fullcflowmodel}
\end{equation}

In general, $\frac{dp}{dT}$ will be set by the local gravitational field and
magnetic pressure.  If the gravitational field is relatively smooth as can be
expected in dark matter haloes, then the pressure will remain nearly constant
in small regions that will begin cooling.  Whether this occurs depends on the
development of thermal instability as discussed in the previous section.  If
$\frac{dp}{dT}$ term is small, this expression simplifies to 

\begin{equation}
\frac{dL_x}{dT_K}=  \frac{5}{2} \frac{\dot{M} k}{\mu m_p}.
\label{eqn:isobaric}
\end{equation}

\noindent
which is known as the standard isobaric cooling flow model.  If the density is
constant which would result from high magnetic pressure, it alternatively reduces to 

\begin{equation}
\frac{dL_x}{dT_K}=  \frac{3}{2} \frac{\dot{M} k}{\mu m_p}
\end{equation}

\noindent
which is for isochoric cooling.
An adiabatic equation would result if cooling were ineffective and the
gravitational potential was strong and the left
hand side would be zero.  In general, $\frac{dp}{dT}$ could be quite
complicated, but note that it is fairly well established observationally that
over the whole cooling flow volume the temperature derivative, $\frac{dT}{dr}$
is positive and the pressure derivative is negative, $\frac{dp}{dr}$.
Therefore, the last term is likely to be positive, and the relative amount of
X-ray luminosity would be emitted according to somewhere in between the
previous two equations if X-ray radiative cooling dominated the energy release.

Note that Equation~\ref{eqn:fullcflowmodel} is just the first law of thermodynamics ($dQ=dU+pdV=\frac{3}{2} N k
dT +d(pV)-Vdp= \frac{5}{2} N k dT - V d p=
\frac{5}{2} N k dT - N \frac{dp}{\rho}$) differentiated with respect to
time ($\frac{5}{2} N k \frac{dT}{dt} - \frac{N}{\rho} \frac{dp}{dt}
kT=\frac{dQ}{dt}$) with X-ray cooling as the only heat loss term.

Equation~\ref{eqn:isobaric} is also frequently expressed in terms of the emission measure, 

\begin{equation}
\frac{dEM}{dT_K} = \frac{5}{2}  \frac{\dot{M} k}{\mu m_p}  \frac{1}{\Lambda}
\end{equation}

\begin{figure}[ht]
\begin{center}
\includegraphics[width=0.9\columnwidth]{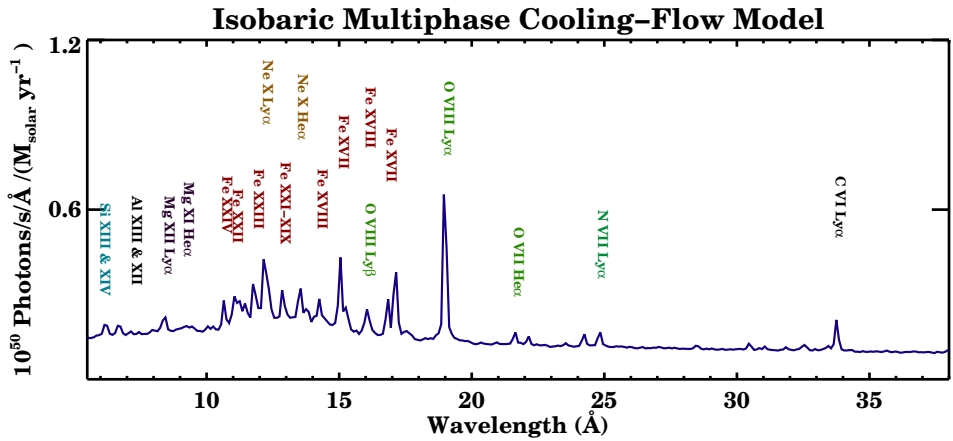}
\end{center}
\caption{\label{fig:cflowmodel}The standard isobaric cooling-flow model.  The model is produced by
summing collisionally-ionized X-ray spectra within a temperature range such
that the relative amount of luminosity per temperature interval is a
constant.  This model predicts relatively prominent iron L shell transitions
between 10 and 17 $\mbox{\AA}$ that arise from a wide range of temperatures.
In particular, Fe XVII which is emitted between 500 eV and 1 keV has strong
emission line blends at 15 and 17 $\mbox{\AA}$.}
\end{figure}

This then can be used in conjunction with the atomic physics necessary
to produce an X-ray spectrum as shown in Figure~\ref{fig:cflowmodel},
which can be compared with the data.  X-ray spectroscopy has
demonstrated that this model is incomplete and therefore we have most
likely neglected additional heating or possibly cooling in our
derivation of this model.  The following section describes the
instrumentation and analysis necessary to test it.

\clearpage
\section{X-ray Instrumentation and Observational Techniques}

A number of X-ray instruments have been launched above the atmosphere,
which is opaque to X-rays, to study X-ray sources in the last 40
years.  X-ray telescopes typically reflect X-rays at grazing incidence
using mirros coated with a high-Z material. CCDs, proportional counters,
and microchannel plates have been used to record the positions of
X-ray photons and make low resolution energy measurements. Crystals, reflection
and transmission gratings have been used to disperse X-rays and
produce high resolution spectra.

Below, we first discuss some of the important characteristics of X-ray
telescopes and the relevant instrumentation that is used to study clusters of
galaxies in X-rays.  Then, we discuss the analysis techniques that are used to
interpret the data collected from these satellites.

\subsection{X-ray Telescopes and their Relevance to Clusters}

X-ray photons from astrophysical sources have been detected by a number of X-ray
telescopes.  For observations of clusters of galaxies, several instrument
characteristics have been important for gaining a complete description of the
X-ray emission.   

{\bf Spectral Resolution:}  First, the spectral resolution or resolving power
($\frac{\lambda}{\Delta\lambda}$ or $\frac{E}{\Delta E}$) affects the ability
to extract useful information from the spectrum.  For typical temperatures in
clusters of galaxies, the Fe L complex most strongly constrains the
distribution of temperatures.  Merely detecting or not detecting an emission
line from a given charge state from a given ion, places narrow constraints on
the distributions of temperatures in the plasma.  The required resolution to
resolve lines from individual Fe L ions is about 100 at 1 keV.  Most X-ray
detectors, which historically have been either proportional counters or
solid-state devices, have been unable to achieve this resolution and therefore
the Fe L complex appears as an unresolved bump in the spectrum.  Proportional
counters work by the incoming X-ray being photoelectrically absorbed an
inert gas atom followed by
the measurement of the electrical discharge induced by the cascade Auger
electrons and the further release of fluorescent X-rays. They typically achieve resolution near
20\%.  CCD devices work by the X-ray
creating a photo-electron in the silicon, which creates a electron-hole pair
cascade through ionization.  The electrons drift to a set of surface contacts
due to an electric field and the number of electrons is used to measure the energy of the
incident X-ray.   These typically have resolving powers near 10 or 20
at 1 keV.  These devices can,
however, measure an average temperature accurately from the shape of the
continuum, which is sensitive to the bremsstrahlung emission, or from the
centroid of the Fe L complex.  These instruments, however, have difficulty
contraining the {\it distribution} of temperatures composing a single spectrum.

  Dispersive instruments, which either use a grating or crystal, have achieved high spectral
resolution in the Fe L band.  Crystals disperse X-rays according to the Bragg
condition in which constructive interference is set up by the X-ray wavelength
being equal to a multiple of the projected crystal spacing.  Reflection and
transmission gratings disperse X-rays by a constructive periodic surface where
the X-ray wave will interfere constructively for a given angle of incidence.
Both the Reflection Grating Spectrometers on XMM-Newton, and the Focal Plane
Crystal Spectrometer on the Einstein Observatory produced high resolution
X-ray spectra of clusters.  Future missions may include non-dispersive
microcalorimeters, which use the temperature change of a cryogenically-cooled
absorber due to the photo-electron to measure the energy of the X-ray.  The
Astro-E1 and Suzaku missions included a microcalorimeter, but unfortunately
the rocket failed for Astro-E1 and the microcalorimeter did not operate long
enough to observe cosmic X-rays for Suzaku.  A microcalorimeter can achieve
sufficiently high spectral resolution in the Fe K band such that few hundred
km/s velocities of the intracluster medium could be measured.

{\bf Effective Area and Exposure Time:} The effective collecting area
and the exposure time of the observation determine the number of
photons collected. The number of photons enters into the effectiveness
of the observation in a number of ways.  Typically, tens of thousands
of photons are collected in a given observation.  Hundred of photons
are necessary to make any detailed surface brightness image.
Thousands of photons are necessary to make detailed spectroscopic
measurements of the temperature.  Tens of thousands of photons are
necessary to make detailed measurements of elemental abundances as
well as the construction of the differential emission measure.

{\bf Field of View:}  Field of view determines the fraction of photons from a
cluster that are detected.  Often, the field of view has been smaller than
the sizes of typical nearby clusters and therefore are an important factor in
comparing results from different instruments.

{\bf Spectral Bandpass:}  The spectral bandpass of most X-ray instruments has
been quite large and for the most part encompassed most of the X-ray spectral
band.  Most of the photons from clusters of galaxies are emitted below 2 keV,
and most instruments have had high efficiency at these energies.  

{\bf Angular Resolution:}  Angular resolution is an important factor for
spatially-resolved spectroscopy.  XMM-Newton (with FWHM of 6'') and Chandra
(with FWHM of 0.5''), for example, have
been able to perform spatially-resolved photometry on kiloparsec scales for
nearby clusters.  

\subsection{Analysis Techniques}

Analysis techniques play a vital role in the interpretation of X-ray spectra
from clusters of galaxies.  The data analysis is usually quite complex
compared to the techniques that can be applied to data from unresolved
sources.  In addition, the instrument response functions are actually quite
complex, and therefore the application of these functions is usually
problematic and results often are somewhat model dependent.

X-ray photons that are recorded in detectors after being reflected and dispersed by
optics have three quantities that are measured:  two detector coordinates,
$x$ and $y$ and one energy measurement, $p$.  These are
indirectly related to the position of the photon on the sky, $\phi$ and
$\psi$, and the photon's intrinsic energy, $e$.  For dispersive spectrometers
the relationship between these three variables and their detector counterparts
are very indirect and the full convolution has to be considered.  For
non-dispersive spectrometers, the relationship is more direct so
approximations can be used.  For example, it is customary to assume that $(x, y)
\approx (\phi, \psi)$ when extracting a spectrum to perform spatially-resolved
spectro-photometry or to assume that $p \approx e$ to construct an image.  

The detection probability, $D$, for a photon emitted from solid angle
position, ${\bf \Omega}$  and energy, $e$ with measured values of detector
coordinates ($x$, $y$) and CCD pulseheight $p$ is given by

\begin{equation}
D (x,y,p) = \int de~d\Omega ~ R ( x,y,p~|~ e,{\bf \Omega} ) ~
\frac{d^2F (e,\Omega)}{d\Omega dE} 
\end{equation}

\noindent where $\frac{d^2F}{d\Omega dE}$ is the spatially-varying
spectral source model and $R$ is the instrument response function.
All existing data analysis methods either solve this integral by, 1)
making approximations about the response function, 2) use a Monte
Carlo approach \cite{peterson04}, or 3) perform an integration over
the solid angle and just
study the spectra 4) or perform an integration over the spectra and
just study the image.  One can easily see some of the reasons for
confusing results on spectra from clusters of galaxies.  It is often
somewhat ill-defined by what one analysis would mean when compared to
an analysis from a different instrument without specifying the full
source function, $\frac{d^2F}{d\Omega dE}$.

  Note that most analyses construct a
response matrix and an ancillary response file, which are used in spectral
codes like XSPEC \cite{arnaud96} or SPEX \cite{kaastra96}.  These codes
multiply a source spectrum on a grid by the response matrix, which relates the
input spectrum to the model observed spectrum.  The model observed spectrum
can be compared to the real data by statistical tests, typically a binned
$\chi^2$ calculation.  The response matrix ($R_m$) /ancillary response ($A$)
file approach is a simplification of the response function above by assuming
the following separation,

\begin{equation}
R(x,y,p~|~e,\phi,\psi) \sim R_M(p~|~e) A(e)
\end{equation}

\noindent where $A$ is constructed separately for each spectrum
depending on the spatial position ($x, y$) of the photons being
studied.  This approach may or may not be appropriate for a particular
analysis.

\noindent The source function is in turn related to the function in
  Equation 7 in \S3.1.5 differentiated with respect to spatial
  coordinates.

\begin{equation}
\frac{d^2F}{d\Omega dE} = \int_0^{\infty}
~dT \frac{d\alpha}{dE} \frac{d^2EM}{d\Omega
  dT}
\end{equation}

\noindent
where $\frac{d^2EM}{d\Omega dT}$ is the differential emission measure per solid
angle.  This function is the true observable if the plasma is
assumed to be in collisional equilibrium.  Note that if a given patch of the
sky is assumed to be isothermal this will directly give a measurement of the
density, since the volume element can be related to the angular coordinate
$\Omega$ for an assumed source distance, assuming some geometry to
obtain the depth.  The temperature structure in clusters is never completely
isothermal, so a variety of temperatures will be sampled along the line of sight.

 We will not review all work in
dealing with these data analysis problems, but refer to the individual work in
the following section for more details since it can vary considerably between authors.

\clearpage
\section{X-ray Spectra of Cooling Clusters}

A wealth of information has been collected by a number of X-ray satellites on
cooling core clusters.  Recently, there have been some strict tests to
various aspects of the cooling-flow scenario.  We review many of the previous
spectroscopic and imaging results that led to the formation of the
cooling-flow model for cluster thermal evolution.  We then discuss how high
resolution X-ray spectroscopic results have questioned a number of assumptions
in the interpretation of the previous work.  These observations have largely
galvanized support around the idea that the thermodynamic evolution of cluster
of galaxies is considerably different than was previously thought.
We then review some recent X-ray imaging and spectro-photometric work that has
left some clues to the resolution of the cooling-flow problem.  Finally in
this section, we define the structure of the cooling-flow problem in clusters
of galaxies.

\subsection{Early Work on Imaging Observations}

The cooling time of the ICM is less than the Hubble time at the
centers of many clustes as demonstrated by \citeextra[Lea et
al.;lea73] with the Uhuru mission. Imaging observations established
the existence of sharp surface brightness peaks in some cluster of
galaxies.  These surface brightness peaks and short cooling times were
interpreted as strong evidence for the existence of cooling flows.
The existence of sharp surface brightness peaks were found by the
Copernicus satellite \cite{fabian74,mitchell75}, rocket missions
\cite{gorenstein77}, and SAS-3 \cite{helmken78}.  Later missions
established the widespread nature of the phenomenon by establishing
that approximately 50\% of all clusters contained sharp surface brightness
profiles.  The Einstein Observatory \cite{stewart84,arnaud88}, EXOSAT
Observatory \cite{lahav89,edge91,edge92}, and ROSAT Observatories
\cite{allen01a} confirmed this picture with observations of hundreds
of nearby clusters of galaxies.

Considerable work was done to establish the implied mass deposition
rates from the cooling flows \cite{jones84}.  The mass deposition
rates, ${\rm \dot{M}}$, were estimated to be between 1-10 solar mass
per year for a giant elliptical galaxy and 100-1000 solar masses per
year for the largest clusters of galaxies.  A first order estimate of
the mass deposition rate is simply made by taking the total gas mass
calculated from the density profile and dividing that by the cooling
time at the edge of the region.  Alternatively, this can be written as
the luminosity of the cooling flow volume divided by the temperature
(in units of kT) times the mean mass per particle.  More detailed
calculations take into account the work done on the gas by its
compression in the gravitational potential
\cite{thomas87,white87a,white87b,white87c,white88,allen93}. Generally,
this effect is thought to reduce the overall level of mass cooling by
at most a factor of two from the amount predicted without
gravitational compression \cite{arnaud88}.

\subsection{Early Work on Low Resolution Spectroscopy}

The modeling of the X-ray spectra of cooling clusters has become
considerably more mature in 30 years.  In addition, some of the early
work with cooling-flow models has now been discounted.  This work
exemplifies the difficulty in applying models at low spectral
resolution.  We briefly discuss some of the major changes to modeling
the X-ray spectrum of cooling clusters.

The Areil-V satellite established that the emission from clusters of
galaxies was thermal (i.e. likely in collisional equilibrium) by
detecting the emission lines from the Fe K$\alpha$ blend (Fe XXV
He$\alpha$, Fe XXVI Ly$\alpha$) from the Perseus cluster
\cite{mitchell76}.  Previously, there were arguments for the origin of
the X-ray emission from clusters as either inverse compton scattering
of microwave background photons off relativistic electrons or emission
from a collection of unresolved X-ray binaries. The detection of
thermal emission, however, established that the hot gas was trapped in
the dark matter potential and emitted at a temperature close to what
was expected from the velocity distribution of member galaxies.

\begin{figure}[ht]
\begin{center}
\includegraphics[height=0.9\columnwidth,angle=-90]{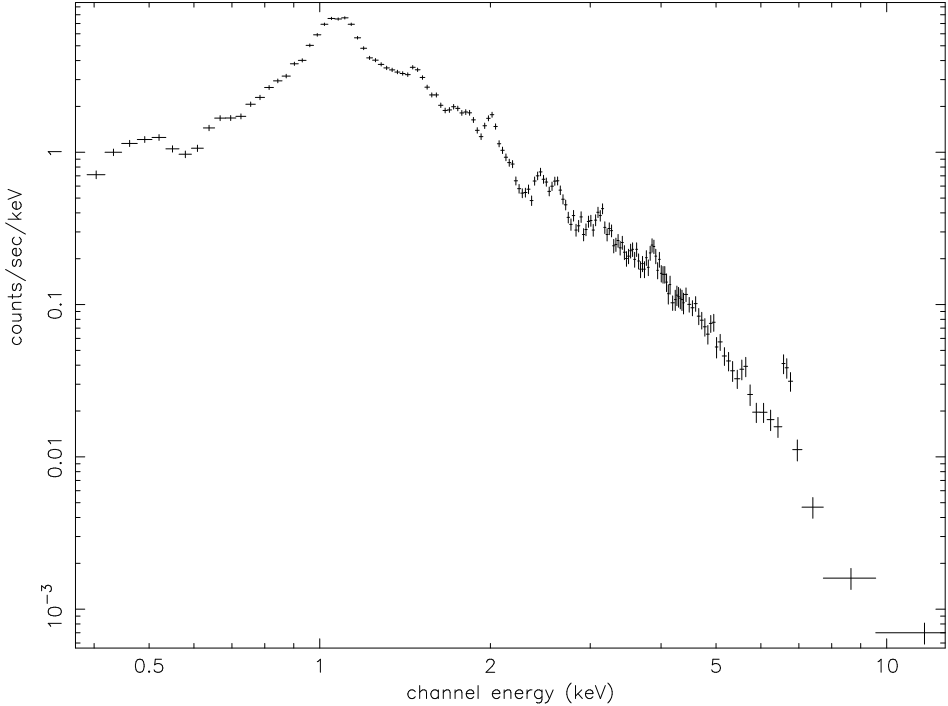}
\end{center}
\caption{The X-ray spectrum of the central region of the Virgo cluster of
galaxies obtained with the ASCA observatory.  This spectrum is one of the
highest quality spectra obtained prior to the launch of the RGS instruments on
XMM-Newton.  The bumps in the spectrum are due to unresolved blends of
emission lines.  In particular, the bump near 1 keV is due to a forest of Fe L
shell transitions, which are quite sensitive to the temperature of the plasma.}
\end{figure}

After the thermal model was established, hundreds, if not thousands, of
clusters had their ``mean'' temperature measured with solid state detectors or
proportional counters most frequently on the Einstein, EXOSAT, Beppo-SAX, and ASCA observatories.  This work established relationships between many observables in
clusters.  In particular, relationships between temperature and luminosity
\cite{david93,white97,markevitch98,wu99,horner99}, mass
\cite{nevalainen00,xu01}, and entropy \cite{ponman99} indicate that mean cluster
gas properties roughly follow from what is expected from self-similar gravitational
collapse.  There may be some deviation from these relations, more
preferentially in lower mass clusters indicating the importance of
non-gravitational processes in cluster formation and evolution.

It was discovered quite early that many clusters do not have an
isothermal structure \cite{ulmer78} and have softer X-ray spectra in
the core.  This implies lower mean temperatures.  A number of more
complicated spectral models were later used to represent the X-ray
emission from cooling-flows with EXOSAT, Einstein, Beppo-SAX, and ASCA
observations.  Two temperature models were used by a number
of authors to represent the distribution of temperatures in the cores
of clusters
\cite{matsumoto96,ikebe97,xu98,fukazawa98,fukazawa00,makishima01}. An
alternative interpretation of the cold gas in the cores of cluster was
developed by \citeextra[Makishima et al.;makishima01] where the dark
matter halo of the cD galaxy is responsible for an unmixed ICM
component.

  Others used the cooling flow model discussed in \S3.3 modified by an
absorber.  The absorber was necessary to reduce the soft X-ray
emission at low energies observed by \citeextra[White et al.;white91].
In retrospect it is clear that these observations were demonstrating a
lack of emission at low energies relative to the cooling flow model,
although this was not explicitly stated.  In particular, the
relatively low spectral resolution made it difficult to distinguish
between models.

The absorber in these models was modelled typically as a single
absorbing screen
\cite{white91,johnstone92,fabian94,buote98,buote99,buote00a,allen00,allen01a}.
The screen was placed between the cluster and the observer, but the
assumption was that this absorption was representing cold gas in the
center of the cluster.  The absorption typically had column densities
near $10^{21} \mbox{cm}^{-2}$, which would have implied large
quantities of cold molecular gas if it was distributed uniformly.
Initial attempts to find this quantity of gas were unsuccessful
\cite{odea98}, although considerable cold material has since been
found by \citeextra[Edge;edge01] and \citeextra[Salome \& Combes;salome04].  Another model used was a cooling
flow model with a single absorption edge.  The absorption edge had an
energy near 0.7 keV, which could have been from ionized Oxygen
possibly from warm ($10^{6}$ K) material \cite{buote00b} or dust
\cite{allen01b}.  Later, high resolution Reflection Grating
Spectrometer (RGS) observations demonstrated that the single edge
model mimicked the effect of lack of emission from colder Fe ions.  It
is interesting to note, in particular, that the fitted energy of the
single absorption edge was slightly lower than the 3s-2p transitions
of Fe XVII, and would therefore optimally appear to absorb cooler
emission from the standard cooling flow model.

\begin{figure}[ht]
\begin{center}
\includegraphics[width=0.9\columnwidth,angle=-90]{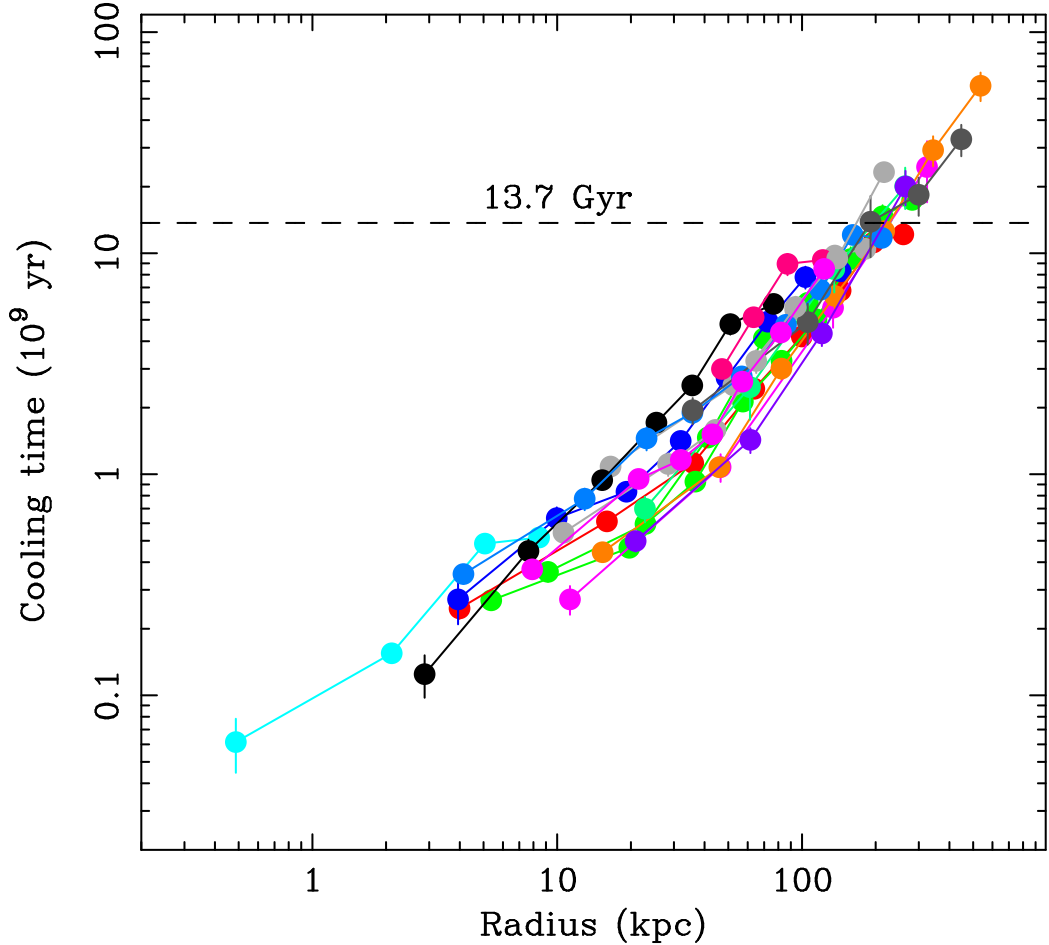}
\end{center}
\caption{The cooling time as a function of radius for a sample of relaxed
clusters as measured with Chandra.  If left undisturbed, all plasma within 100
to 200 kpc would have sufficient time to cool.  Figure adapted from
\citeextra[Voigt et al.;voigt04]. \label{fig:tcool}}
\end{figure}

\begin{figure}[ht]
\begin{center}
\includegraphics[width=0.9\columnwidth,angle=-90]{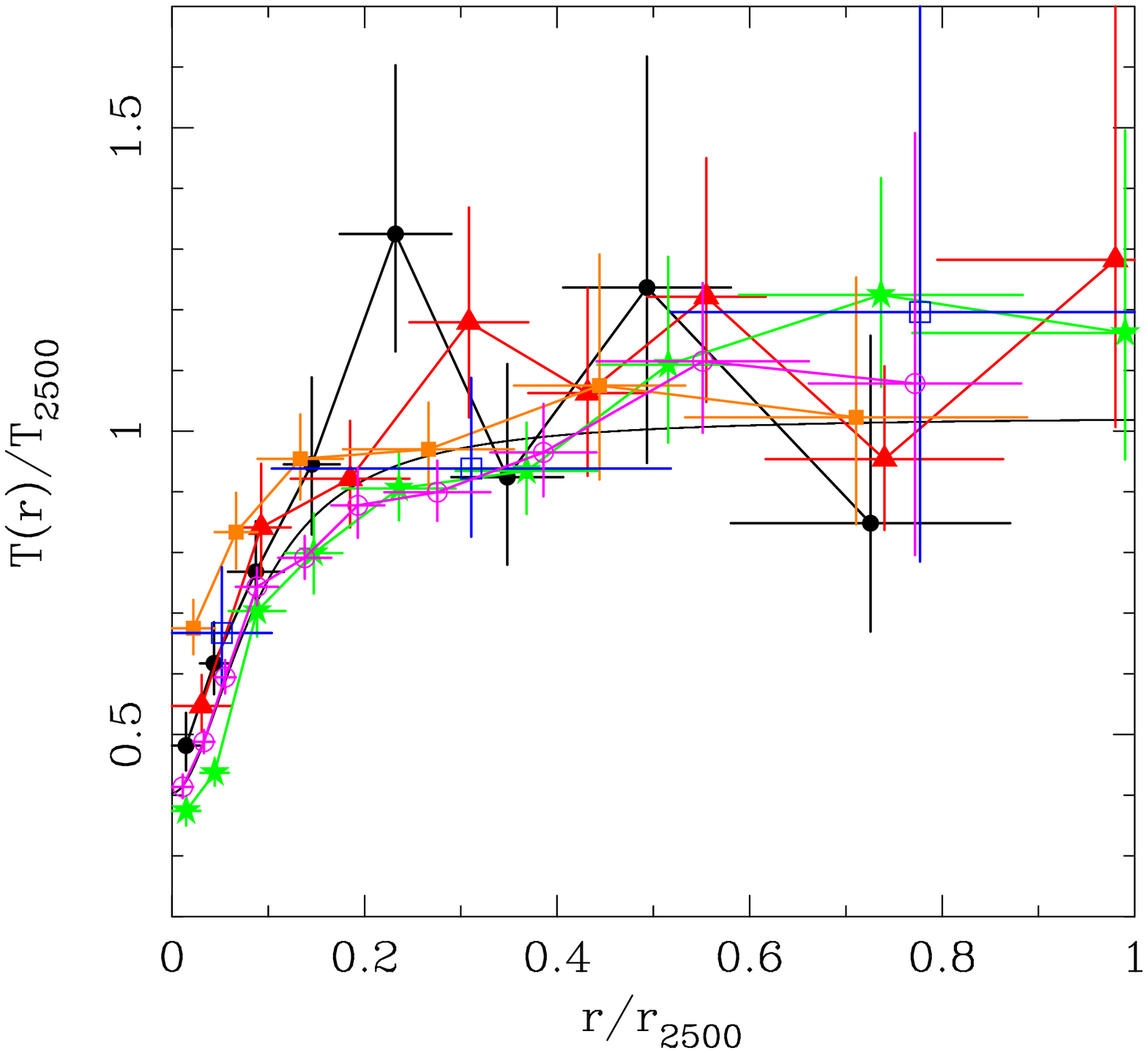}
\end{center}
\caption{The average temperature in radial bins for a sample of relaxed
clusters of galaxies as measured with Chandra.  The temperature and radius are
scaled to $r_{2500}$.  A typical cluster shows a clear decline in the average
temperature at the center, which is in agreement with many spectroscopic
studies over the previous two decades.  Figure is adapted from
\cite{allen01c}. \label{fig:tvsr}}
\end{figure}

Many observations generally established the cooling time is a fraction of a
Gyr in relaxed clusters.  They also clearly indicated that the temperature was
lower where the central surface brightness was highest.  These two fact are
now shown clearly with recent data from the Chandra observatory in
Figure~\ref{fig:tcool} and \ref{fig:tvsr}.

\subsection{Focal Plane Crystal Spectrometer}

The first high resolution observations of cooling clusters comes from the
Focal Plane Crystal Spectrometer (FPCS) on the Einstein observatory.  This
instrument used a crystal to disperse X-rays according to the Bragg
condition.  It therefore could produce a high resolution spectrum by scanning
in angle and therefore scanning in wavelength.  This worked for an extended
source like a cluster as well.

Unfortunately, there were only a limited number of these observations
and the count rates were not particularly high.  These observations
demonstrated the existence of line emission from 8 ions
\cite{canizares79,canizares82} in four different clusters. Some of
these observations are contradicted by Reflection Grating Spectrometer
observations as will be described in the next section, so it is likely
that these observations were partially compromised by high background
rates and, in some cases, relatively few photon counts.

\subsection{Reflection Grating Spectrometer Observations}

Observations made at high spectral resolution with the Reflection Grating
Spectrometers on XMM-Newton have greatly clarified the observational
interpretation of the soft X-ray spectra of cooling clusters.  Although some
information can be gained from the exact shape of the Fe L complex with low
spectral resolution spectrometers, RGS observations resolve emission lines
from each Fe L charge state (Fe XXIV through Fe XVII) and therefore can place
crucial constraints of the amount of gas present between temperatures of 0.4
and 4 keV. 

\begin{figure}[ht]
\begin{center}
\includegraphics[width=0.9\columnwidth]{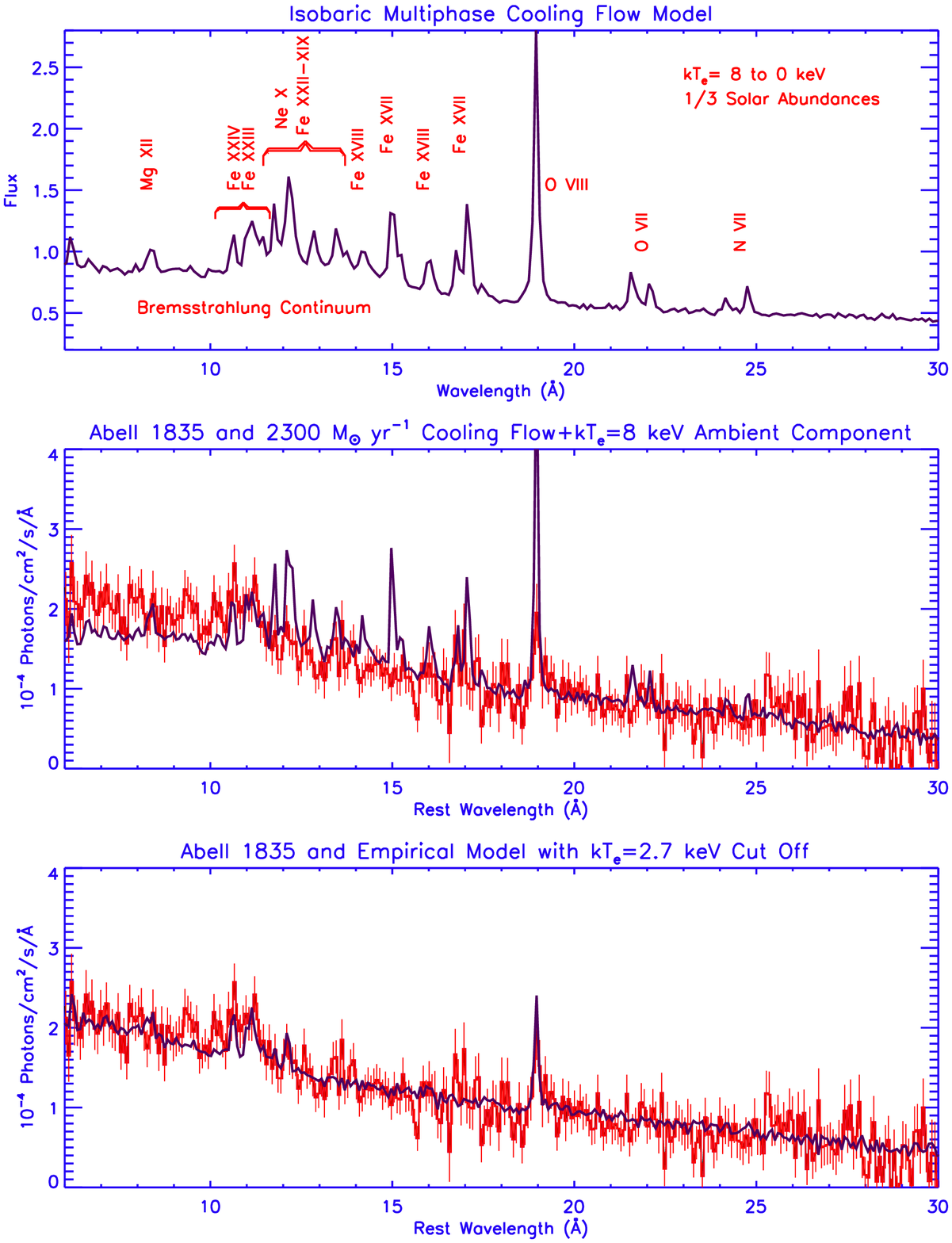}
\end{center}
\caption{\label{fig:1835} Three panels showing the cooling-flow model applied to the X-ray
spectrum of the putative massive cooling-flow, Abell 1835.   The top panel
shows the standard isobaric cooling flow model (see the previous figure).  The
middle panel shows the model (blue) and the data obtained from the Reflection
Grating Spectrometer on XMM-Newton (red).  The model is clearly inconsistent
with the X-ray spectrum, particularly in the prediction of Fe XVII emission
line blends at 15 and 17 $\mbox{\AA}$.  The bottom panel shows the cooling
flow model compared with the data, except all emission coming from
temperatures below 2.7 keV is suppressed.  The explanation for the success of
this model is not known.  Adapted from \citeextra[Peterson et
al.;peterson01].  The spectrum is taken from a 5 by 20 arcminute region of the
core.}
\end{figure}

\begin{figure}[ht]
\begin{center}
\includegraphics[width=0.9\columnwidth]{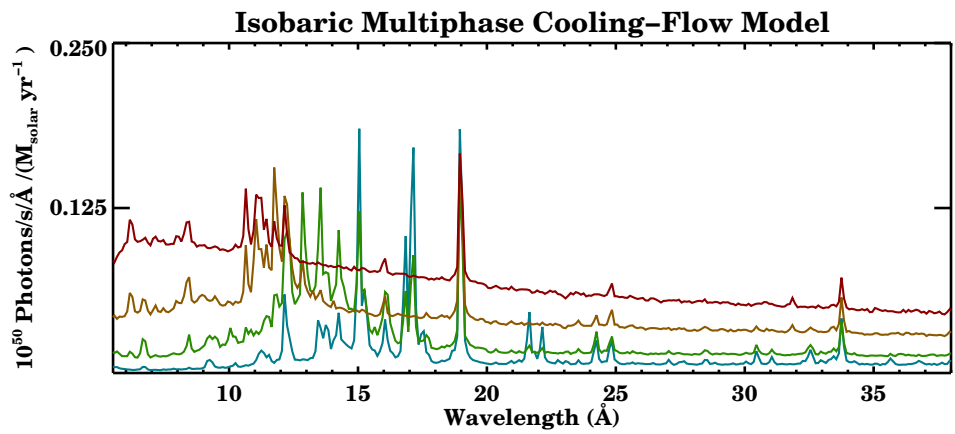}
\end{center}
\caption{\label{fig:4range} The isobaric cooling flow model divided into
the contributions from various temperature ranges, 0.375 to 0.75 keV (blue),
0.75 to 1.5 keV (green), 1.5 to 3 keV (yellow), 3 to 6 keV (red).}
\end{figure}

\begin{figure}[ht]
\begin{center}
\includegraphics[width=0.9\columnwidth]{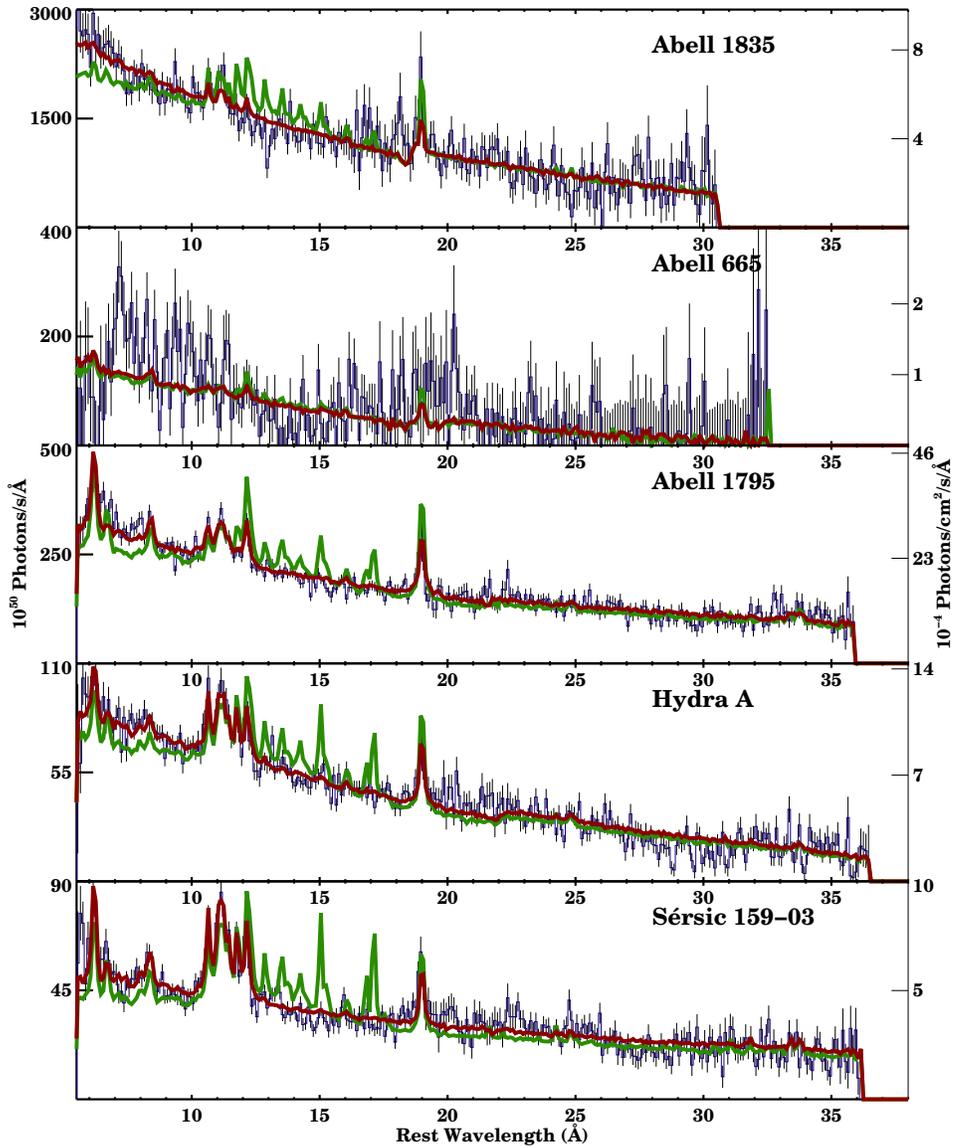}
\end{center}
\caption{\label{fig:spec1} Two models compared with the Reflection Grating Spectrometer data of
a sub-sample of massive cooling flows.  The blue histogram is the RGS data,
the green curve is the standard cooling-flow model, and the red is an
empirical model where emission from the standard cooling-flow model is allowed
to be adjusted in specific temperature ranges. The spectrum is taken from a 5 by 20 arcminute region of the
core.}
\end{figure}
\begin{figure}[ht]
\begin{center}
\includegraphics[width=0.9\columnwidth]{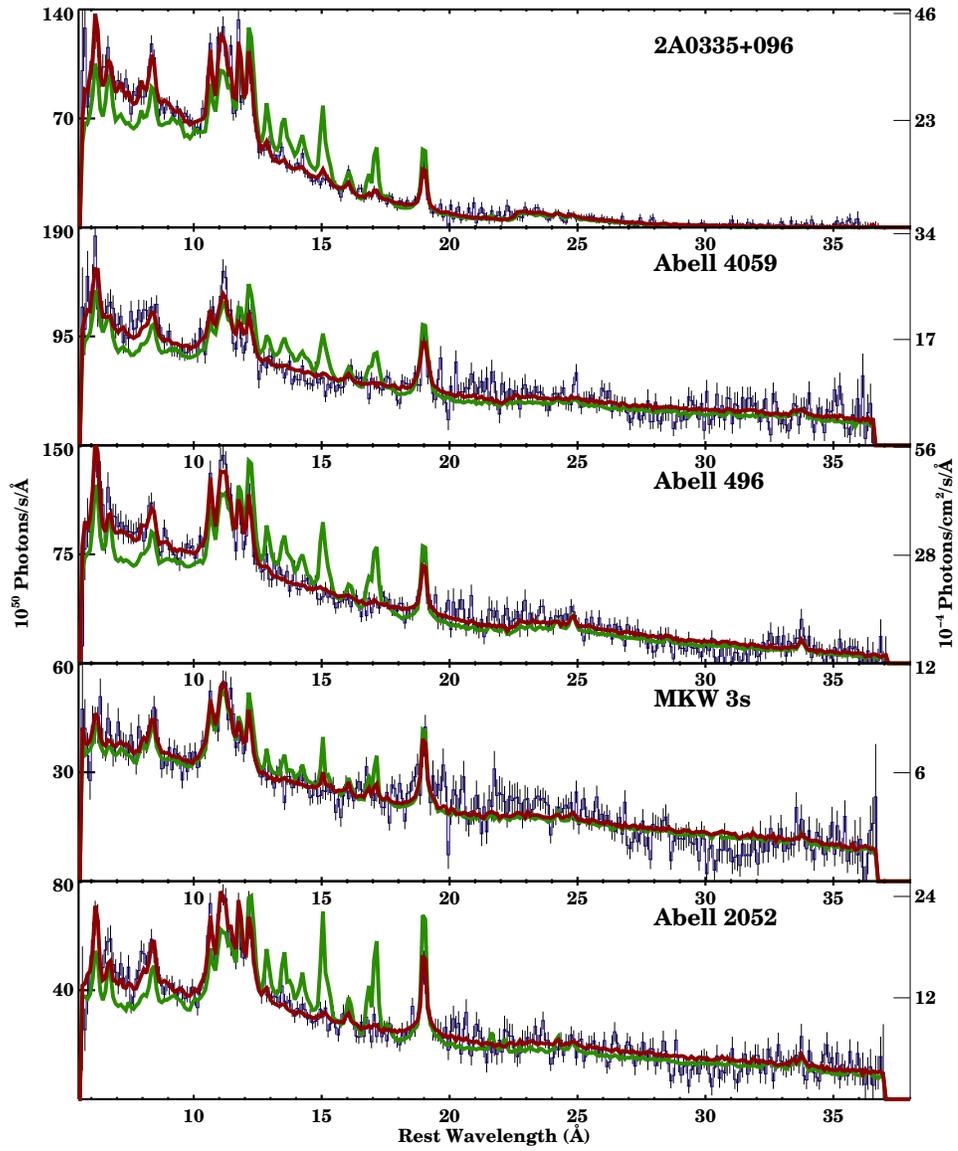}
\end{center}
\caption{\label{fig:spec2}The same as the previous figure, but for a sample of intermediate mass clusters.}
\end{figure}
\begin{figure}[ht]
\begin{center}
\includegraphics[width=0.9\columnwidth]{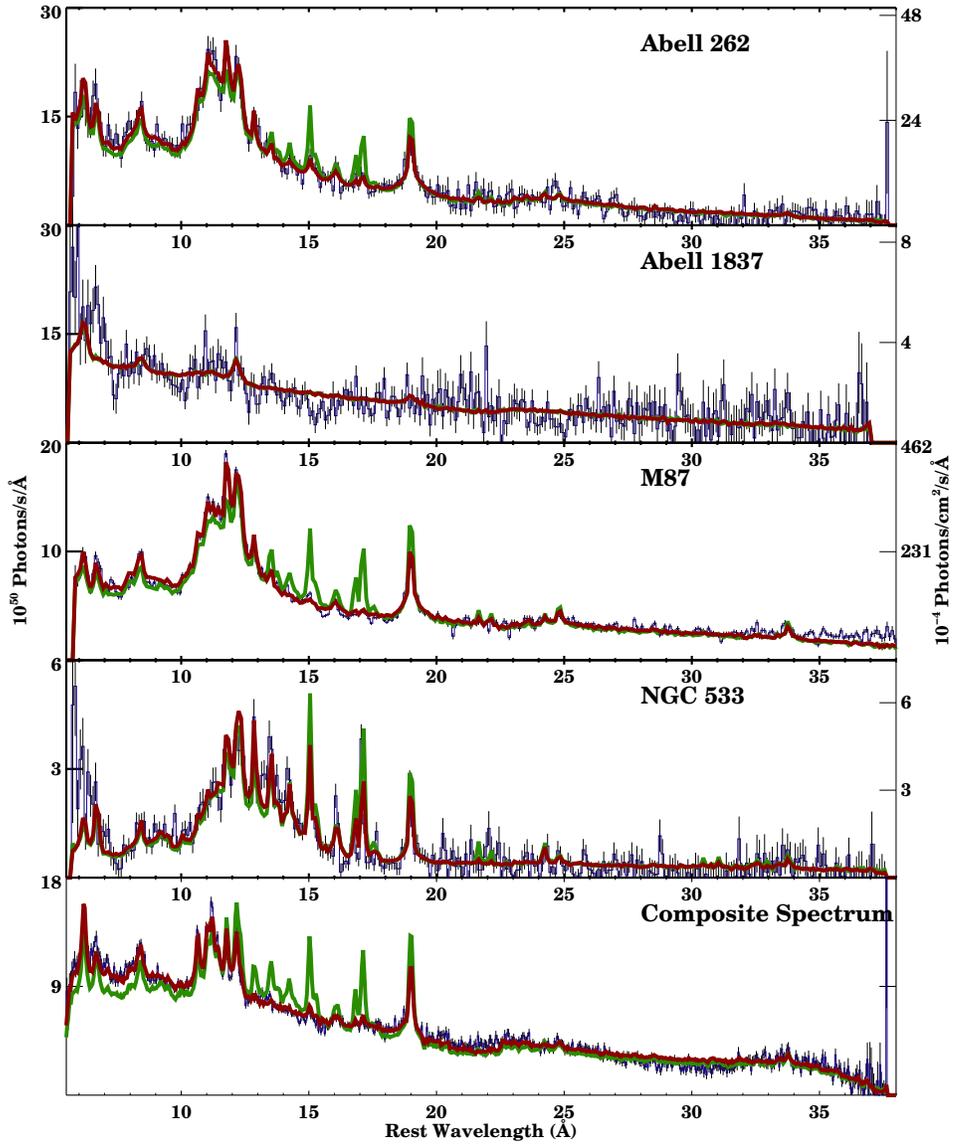}
\end{center}
\caption{\label{fig:spec3}The same as the previous two figures, but a sample of low mass
clusters and groups of galaxies.}
\end{figure}
\begin{figure}[ht]
\begin{center}
\includegraphics[width=0.9\columnwidth]{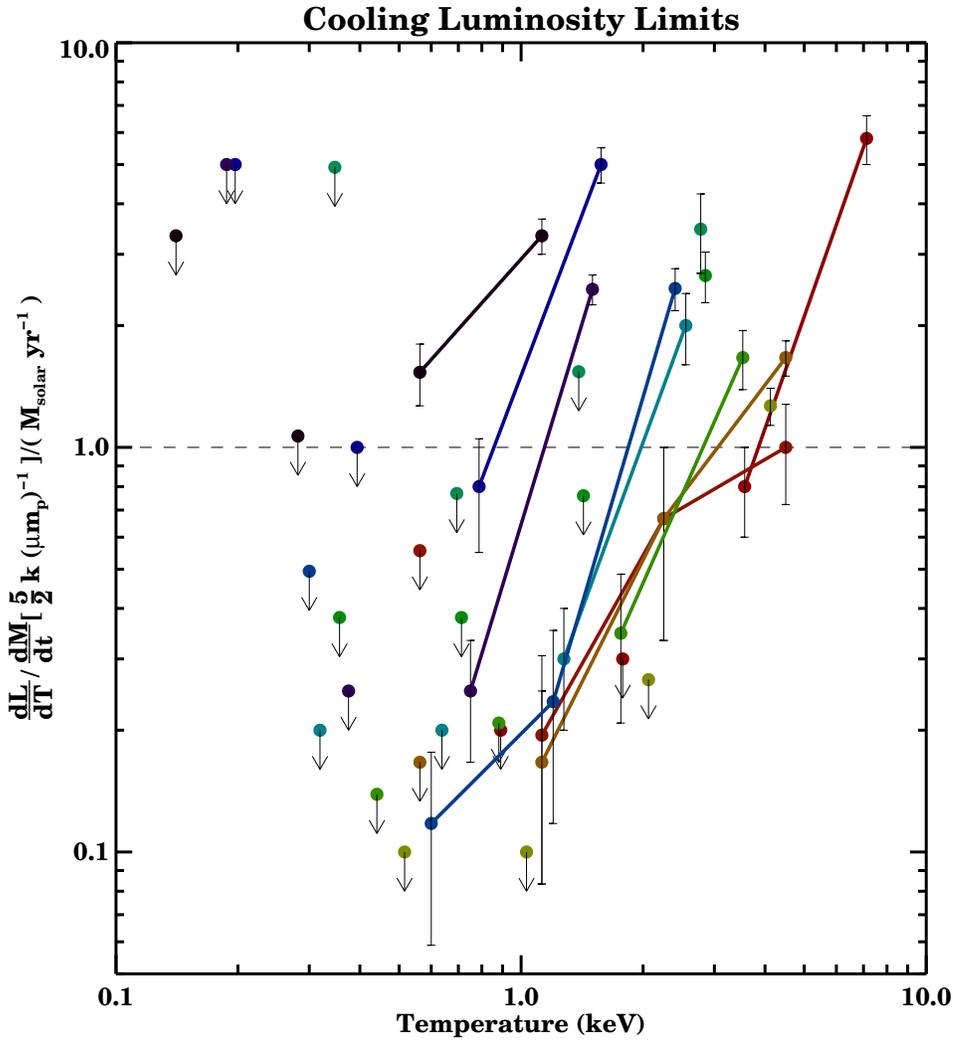}
\end{center}
\caption{\label{fig:dldt1}The relative amount of emission in various temperature ranges from a
sample of 14 clusters obtained with the Reflection Grating Spectrometer on
XMM-Newton.  Each cluster is a different color, and each cluster has four
points.  Points that are not upper limits are connected by a straight line.
The standard cooling flow model predicts the same amount of emission in each
temperature range or a horizontal line.  The data, however, are clearly
inconsistent with that model.  The detected emission has a much steeper
distribution in temperature and many upper limits are a factor of several
below the prediction.  Adapted from \citeextra[Peterson et al.;peterson03].}
\end{figure}
\begin{figure}[ht]
\begin{center}
\includegraphics[width=0.9\columnwidth]{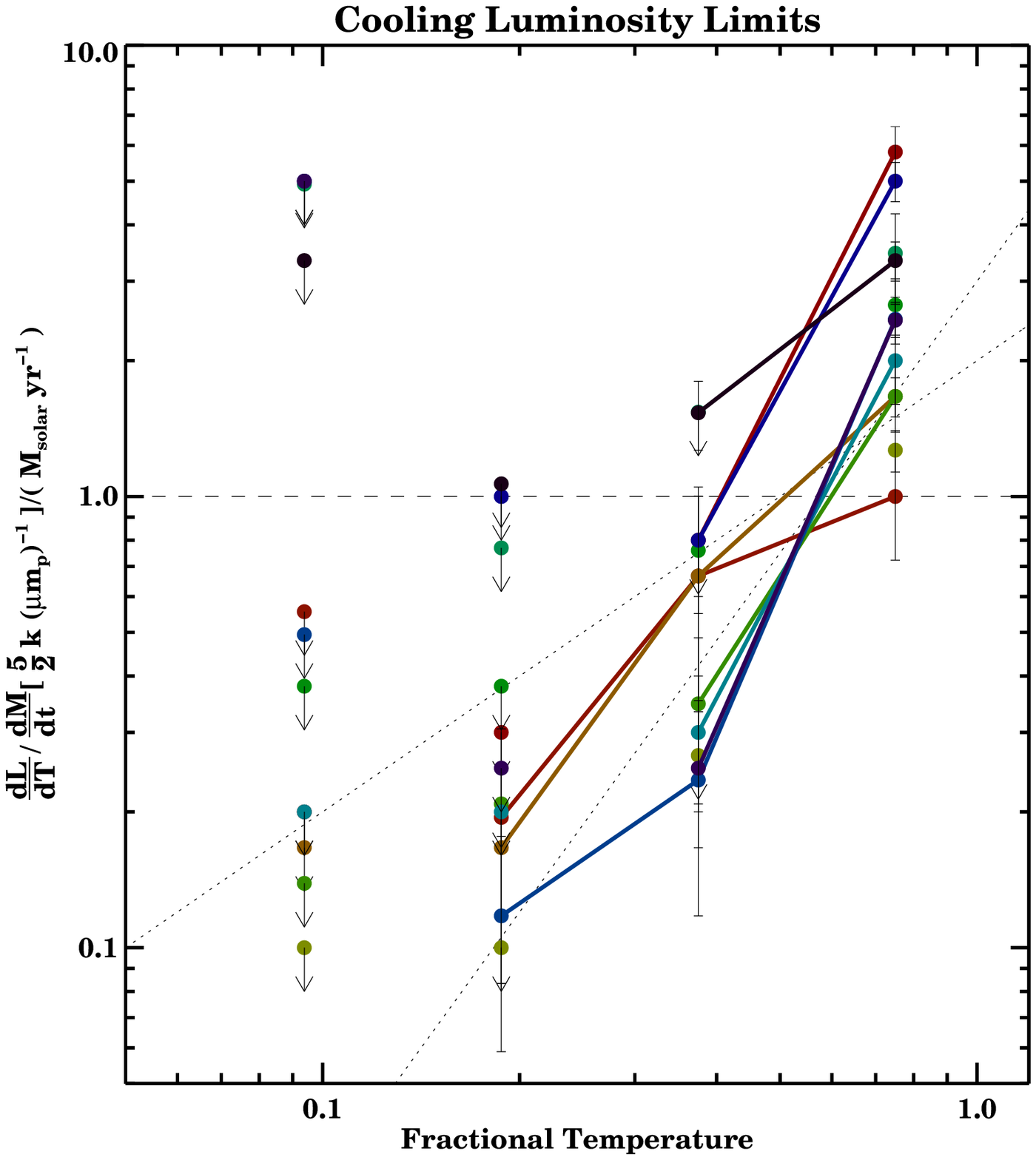}
\end{center}
\caption{\label{fig:dldt2}Same as the previous figure, except the points are normalized to the
maximum temperature in the fit.  A more systematic trend is seen where the
emission follows a linear or quadratic form in temperature (dotted lines).
There is significant scatter in the points, but there does appear to be a
``self-similar'' violation to the standard cooling flow model (dashed line).
Adapted from \citeextra[Peterson et al.;peterson03].}
\end{figure}

The initial application of the standard cooling flow model to a high resolution
X-ray spectrum is shown in Figure~\ref{fig:1835} \cite{peterson01}.  The top panel shows
the spectral prediction between 5 and 35 $\mbox{\AA}$.  Emission lines are expected from Fe L
ions in roughly equal strength between 10 and 18.  Particularly prominent are
emission lines from Fe XVII and 15 and 17 Angstroms.  These emission lines are
produced primarily from plasma between 300 and 700 eV and represent the last
major emitting ion before a cooling plasma would cool to sub-X-ray
temperatures.   

High resolution spectroscopic evidence for lack of cool gas has been
documented in a number of clusters
\cite{peterson01,tamura01,kaastra01,sakelliou02,tamura01b}.  Characteristically, gas
appears to be missing at near a third of the maximum temperature.  This
phenomena has also been documented in elliptical galaxies.   \citeextra[Xu et al.;xu02] found
that Fe XVII and Fe XVIII was present in the nearby elliptical galaxy, NGC
4636.  O VII, however, was not detected and has not been detected in any
galaxy, group, or cluster core.  

A more detailed analysis of the nature of the cooling flow problem is found in
\citeextra[Peterson et al.;peterson03].  A sample of 14 clusters were analyzed in a uniform way to
demonstrate that the cooling flow problem manifests itself at a fraction of
the maximum temperature in the center.  In this sample it is shown that hot
clusters (5-10 keV) generally only show evidence of Fe XXIV-XXII emission and
no other Fe L charge states.  Intermediate temperature clusters (2-5 keV) show
evidence for Fe XXIV-Fe XIX emission, but no Fe XVII and Fe XVIII.  The
coolest clusters and groups (less than 2 keV) show evidence of the entire Fe L
series but anomalously low levels of Fe XVII emission when compared with the
standard cooling flow model.  All of these observations are in contradiction
with the standard cooling flow model, and suggest qualitatively that the model
is violated at characteristically a fraction of the temperature and is more
inconsistent with the model at lower temperatures.

In this quantitative analysis \cite{peterson03}, the cooling flow model was separated
into four different temperature ranges as shown in Figure~\ref{fig:4range}.  Then the
normalization of each temperature range was adjusted to empirically fit the
spectrum shown in Figures~\ref{fig:spec1}, \ref{fig:spec2}, and \ref{fig:spec3}.  The results from this study are shown in Figures~\ref{fig:dldt1}
and \ref{fig:dldt2}.
These graphs plot each of the four luminosity detections for each clusters.
The lines connect the points that are actual spectral detections and not just
upper limits.  The standard cooling flow model would be if all the points
would lie on a horizontal line.  A comparison of Figure \ref{fig:dldt1} and \ref{fig:dldt2} suggests
that when plotting the differential luminosity vs. the fraction of the
temperature as in Figure \ref{fig:dldt2}, that the trend is more systematic.  Therefore,
the emission of a cooling flow is consistent with a power law in differential
luminosity with exponent between 1 and 2.  The cooling flow model, however, is
strongly ruled out in many systems.

At this point, there has been no significant evidence for any significant
difference between cooling flows of similar cluster mass.  This poses some
challenges for heating models where there is a time-dependence to the process.  There also has been no evidence from X-ray
spectroscopy that there are significant quantities of plasma at low X-ray
temperatures above the expectation of the cooling flow model.  In other words,
Figure \ref{fig:dldt2} shows a monotonically decreasing violation of the cooling flow
model and no evidence for gas piling up at some intermediate temperature.
This presents some challenges for continuous heating model that do not reheat
the gas completely.  For both of these reasons, it is not straight-forward to
interpret these results in the context of any heating model 
regardless of the mechanism.

Finally, it is worth noting that these results generally are displaying the
X-ray spectrum of the entire cooling flow volume.  There is considerable work
that is continuing to actually study the changes in the emission spectrum as a
function of spatial position inside the cooling flow.  These studies are
difficult since determining a spatially-resolve differential emission measure
require both high angular resolution and high spectral resolution.  Following,
we discuss some attempts to study the spatial-dependence of the X-ray emission
of cooling flow at moderate spectral resolution.

\subsection{Recent Spatially-resolved Spectro-photometric Observations}

The RGS results obtained at high spectral resolution have been
augmented with further studies at low resolution that attempt to study
the spatial distribution in greater detail
\cite{david01,boehringer01,molendi01,schmidt01,ettori02,johnstone02}.
In general, it has also been demonstrated that the low spectral
resolution observations obtained with XMM-EPIC and Chandra, which are
far more numerous, are generally consistent with the RGS observations.
Significantly smaller mass deposition rates have been measured when
the cooling flow model has been applied.  Similarly, cut-off cooling
flow models have been shown to be consistent with the data.  Needless
to say, a larger range of models can be applied to the low resolution
data and still be statistically consistent with the data.  In general,
it appears that there is no strong evidence for any significant amount
of cold gas (below $\frac{1}{3}$ of the maximum temperature) in any
cluster.

The radial profile of the temperature distribution in cooling flows is
currently in dispute.  It is clear that the ICM has a much narrower range of
temperatures than compared to the standard cooling flow model as established
by the RGS observations.  It is not so clear, however, how narrow that
distribution is at any given radius.  \citeextra[Molendi \& Pizzolato;molendi01] have argued that the ICM
is nearly isothermal at a given radius and any departure from this is due
solely to azimuthal variations.  \citeextra[Buote et al.;buote03] and \citeextra[Kaastra et al.;kaastra04] have argued, however, that nearly a factor of two range of gas
temperatures exists at any radius.

X-ray cavities have been discovered in a number of X-ray imaging
observations. The cavities are presumably excavated by cosmic rays
produced in outbursts by the central radio source \cite{mcnamara01}.
These cavities are important in the study of cooling flows for a
number of reasons.  First, they represent direct empirical evidence
that the cores of cluster have been disturbed.  The work done by
expansion of the bubbles assuming all the energy is deposited in the
cooling flow is within a factor a few to the value required
\cite{birzan04}.  Second, they are relatively intact and coherent
structures indicating the level of suppression of thermal transport
processes possibly by a magnetic field.  Therefore, they provide a
source of external energy to the ICM and detailed study will
eventually give us a complete picture of the transport of that energy.

Constant pressure surface brightness discontinuities, or ``cold
fronts'', are present in many clusters of galaxies
\cite{markevitch00}.  Cold fronts show that there are vast quantities
of plasma oscillating with respect to the gravitational potential.
Some may be cooler, denser groups that have fallen to the center,
whereas others may be cooled gas that already was present in the core
of the cluster but is oscillating because of a merger in the cluster
outskirts \cite{churazov03}.

\begin{figure}[ht]
\begin{center}
\includegraphics[width=0.9\columnwidth]{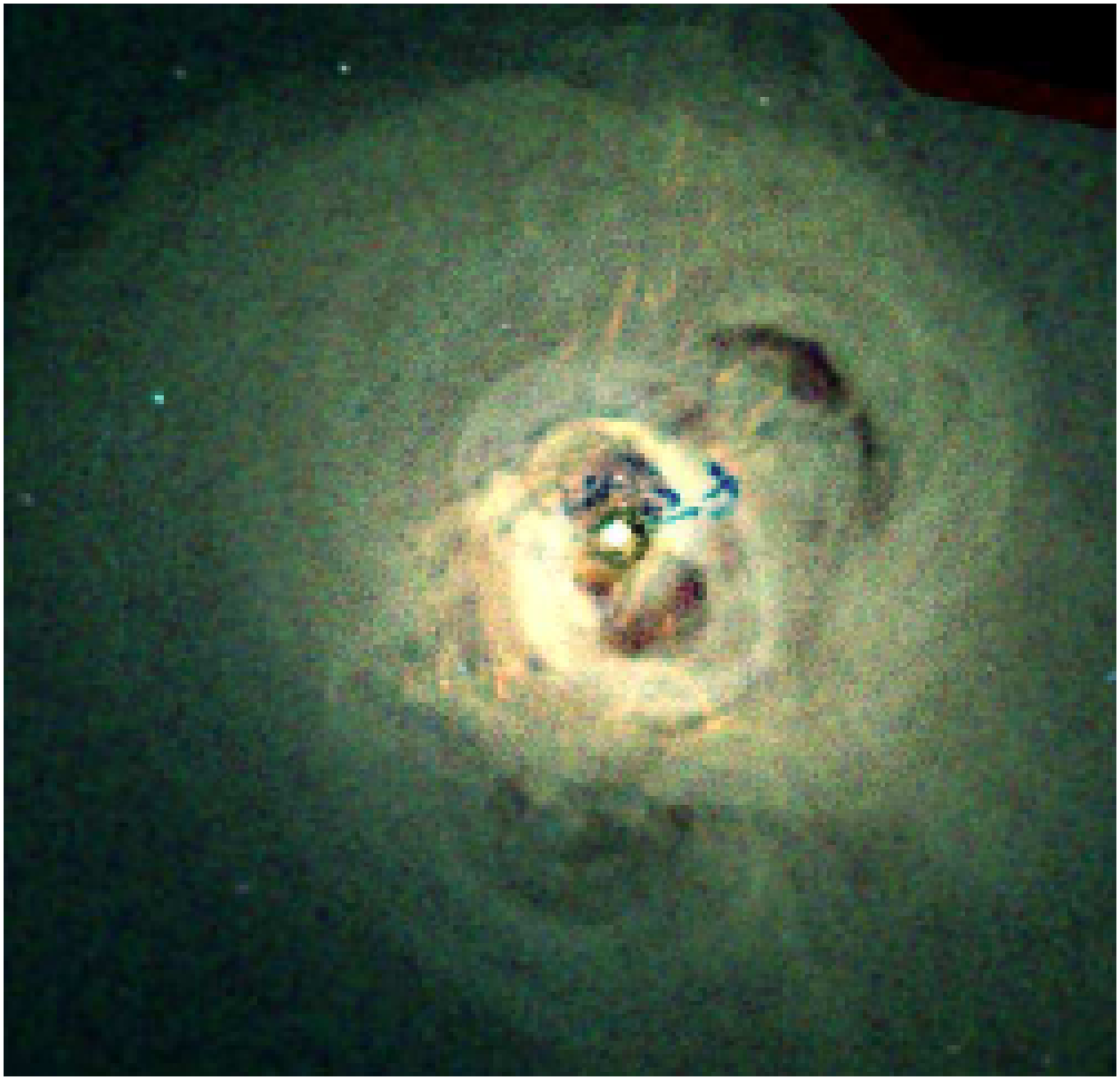}
\end{center}
\caption{High resolution X-ray image of the Perseus cluster of galaxies.  The
bright source in the center is the active galactic nucleus of NGC 1275.  The
two adjacent holes in the X-ray emission as well as the mushroom shaped
depression in the upper right are believed to be cavities in the ICM that have
been excavated by cosmic rays expelled from the active galactic nucleus.
Figure is adapted from \citeextra[Fabian et al. 2005;fabian05b].}
\end{figure}

\begin{figure}[ht]
\begin{center}
\includegraphics[width=0.9\columnwidth]{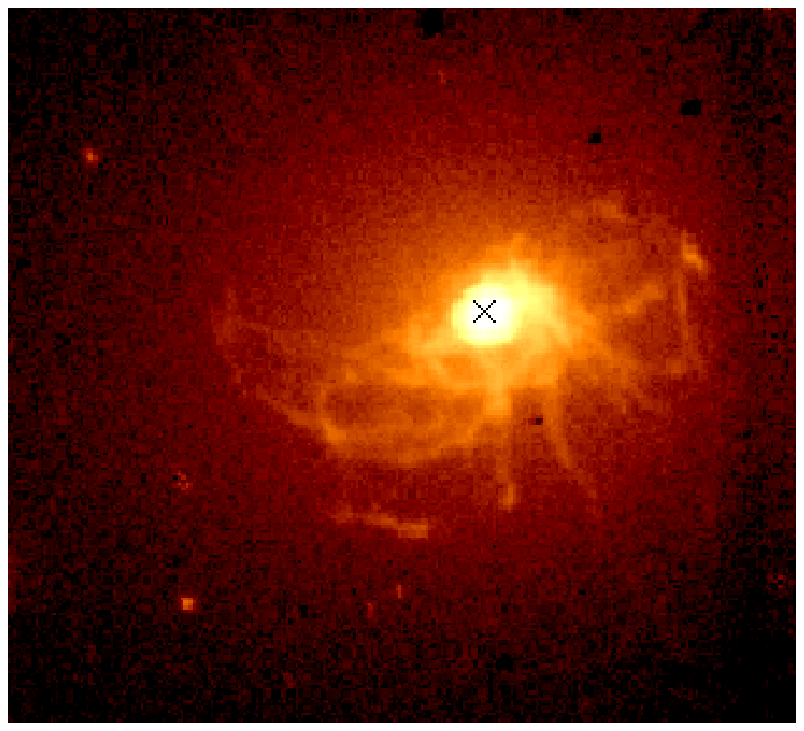}
\end{center}
\caption{Image of the Centaurus cluster of galaxies overlayed with H$\alpha$
emission showing a filamentary structure.  Figure is adapted
from \citeextra[Crawford et al.;crawford05].}
\end{figure}

Fine ripples have been identified in the cluster gas of Perseus
\cite{fabian03a}.  These features are extremely weak and more study is
needed to measure the temperatures across the discontinuities and
confirm their properties. Detections of these features have also been
claimed in Hydra A \cite{nulsen04} and M 87 \cite{forman03}.  It is
unclear if these are an entirely separate phenomena from cold fronts.
Given their proximity to the center, \citeextra[Fabian et
al.;fabian03a] suggested that these were weak shocks and sound waves.
In addition, \citeextra[Sanders et al.;sanders04] have identified an
asymmetric ``swirl'' in the temperature distribution in the Perseus
cluster. This observational feature is may be due to the effect of
off-center cluster mergers.

\subsection{Observations of Cataclysmic Variables}

Although not the subject of this review, it is certainly worth noting
that the X-ray spectrum of some binary accreting white dwarf systems,
cataclysmic variables, actually resembles the X-ray spectrum of the
standard cooling flow model \cite{mukai03}.  The flow behind the
standing shock in an accreting white dwarf involves the same physical
processes as those expected for a cooling flow in a cluster although
the geometry may be different.  Although there are a great number of
differences between clusters of galaxies and cataclysmic variables,
these observations can be used to confidently say there is no mistake
in applying a model for a radiative-cooling dominated plasma.
Furthermore, this points to no major difficulties in the atomic physics
models or subtle problem with the plasma physics arguments we have
used.

\subsection{Definition of the Cooling Flow Problem}

We now briefly discuss what we believe the cooling flow problem is, and how it
might be resolved.  Clearly, the problem is quite complex and it is difficult
not to see the problem in historical terms.  \citeextra[Peterson et al.;peterson03], for example,
discussed a difference between the {\it soft X-ray cooling-flow problem} and
the {\it mass sink cooling-flow problem}.  The former refers to the recent
discrepancy seen in the soft X-ray spectrum between what was predicted and
what was observed.  The latter refers to the difficulty in detecting any
by-products in cooling clusters from the hypothesized cooling-flow plasma.

These definitions, however, might just categorize our ignorance of the
solution to the problem.  The major difficulty is that: 1) the cluster
plasma loses energy by emitting the very X-rays we detect, 2)
efficient and distributed heat sources are difficult to construct, 3)
the cluster plasma appears to cool {\it most} of the way, but 4)
evidence for complete cooling is utterly lacking.  The cooling-flow
problem as we see it is to understand what happens in the middle of
that process.   After examining whether cooling flows are ruled out, we discuss
many ideas that might alleviate the cooling-flow problem.

Note, from Fig.~6, that time variability on intervals longer than
$10^8~{\rm yr}$ cannot be the solution.

\clearpage

\section{Are cooling flows ruled out?}

In summary, problems with the simple cooling flow picture emerged from two
fronts. The first is the enormous implied mass deposition in some
objects where for example $\dot{M} \sim 1000~{\rm M_{\odot}~yr^{-1}}$ (e.g.
PKS\,0745-191; \citett[fabian85]) which is then not seen in cooled form (the
'mass sink' problem). The central galaxy of this, and many other
brightest cluster galaxies where the surrounding gas have $t_{\rm
cool}<3$~Gyr, has much excess blue light, optical/UV/IR emission-line
nebulosity (\citett[crawford99] and refs. therein) and even molecular
gas \cite{edge01}. The total star formation rate and gas mass is often
one to two orders of magnitude below such a large value of $\dot{M}$
operating for several Gyr. The flow could be somewhat intermittent but
not overly so or short central cooling times would not be so common
\cite{peres98}.  A non-standard IMF for the stars is one
possibility \cite{fabian82}, but has to be extreme \cite{prestwich97}.

The second problem is that a simple cooling flow spectrum is a poor
fit to the data in the soft X-ray band \cite{white91,johnstone92,fabian84}. What was usually done was to fit the
model spectrum of an isobaric cooling flow \cite{johnstone92} plus
an isothermal spectrum to represent the luminosity from gravitational
work done, to the data from the inner regions. A deficit in soft
X-ray emission was found in the data when compared with the model.
This was then modelled as photoelectric absorption due to an absorber
intrinsic to the central regions. The lack of any obvious absorption
in very soft ROSAT spectra was attributed to emission from the
intervening gas in the cluster \cite{allen97}.  

Photoelectric absorption intrinsic to the cluster does not however
appear in the spectra of the jet in M87 \cite{boehringer02} or the
nucleus of NGC 1275 \cite{churazov03}. Nor is it apparent in the
detailed XMM/RGS spectra of cool cluster cores
\cite{peterson01,tamura01,kaastra01,sakelliou02,peterson03}.
Absorption has generally been abandoned as an explanation, although no
detailed test has been made of a model in which the absorption is
intimately linked with the coolest cooling gas clouds. Note though
that the presence of warm emission-line nebulosities, dust-lanes and
clear optical absorption lines \cite{carter97,sparks97} show that at
some level there {\it must} be distributed intrinsic absorption.

Curiously, the 'missing' soft X-ray luminosity (i.e. the above soft
X-ray deficit) is close to the luminosity emitted from the
emission-line nebulosity commonly found in these systems \cite{fabian02a,soker04}. This has led to the suggestion that the gas
{\it is} cooling but not radiatively once its temperature has dropped
to $\sim 1~{\rm keV}$ or so. It can be cooled by mixing \cite{fabian02a}
or conduction \cite{soker04} with cold gas. A possible picture
would be that the gas cools from 100 to within 20~kpc radiatively as a
single-phase flow by which radius its temperature has dropped by a few. It
then shares its temperature with cooler gas and thereby rapidly drops
down to below $10^6~{\rm K}$. The cooler gas is close to the peak of the
cosmic cooling curve and radiates the energy in the UV and optical
bands (and IR if there is molecular gas and dust). Such a solution to
the soft X-ray deficit still suffers from the mass sink problem.

A further situation which could give a soft X-ray deficit is if 
the metals in the ICM are highly inhomogeneous. They are presumably
introduced in a very localized manner and at high abundance by stars
and supernovae. If they do not mix but after time just reside at a
similar temperature to their surroundings then when bremsstrahlung
cooling dominates (which it does above about 1~keV for Solar
abundances) then all the gas cools together. But when it has cooled so
that line cooling dominates, the highly enriched clumps would then cool
rapidly and drop out \cite{fabian01,morris03}. This
has been considered in part by \citeextra[B\"ohringer;boehringer02], although no
conclusive test has been performed.

The main reason that complete cooling flows, in which gas cools from
the cluster virial temperature to well below $10^6~{\rm K}$, are often now
considered to be ruled out is the high spectral resolution XMM/RGS
data \cite{peterson01,tamura01,kaastra01,sakelliou02,peterson03}.
Chandra spectra lead to a similar conclusion \cite{david01,allen01a}.
These spectra have been covered in depth earlier in this review. We
summarize here to note that most data are consistent with the ICM
temperature dropping within the cooling radius by a factor of about 3
(sometimes significantly larger values are found e.g. A2597 \cite{morris05}
and Centaurus \cite{fabian05a}. Data from all wavelengths are
consistent with a residual flow ranging from a few $~{\rm M_{\odot}~yr^{-1}}$ to atypical
value of tens $~{\rm M_{\odot}~yr^{-1}}$ up to $\sim 100 ~{\rm M_{\odot}~yr^{-1}}$ in some cases (e.g.
\citeextra[Bayer-Kim et al.;bayerkim02]; \citeextra[Wise et al.;wise04];
\citeextra[McNamara et al.;mcnamara04]). These values are generally
compatible with the ongoing star formation seen.

\subsection{Cooled gas and star formation}

The brightest galaxy in X-ray peaked clusters often has excess blue
light indicating massive star formation and IR/optical/UV emission
lines \cite{crawford99,donahue00,hicks05}. The regions are often dusty
so the determination of the star formation rate depends on uncertin
dust corrections. It can be tens $~{\rm M_{\odot}~yr^{-1}}$ to about 100~$~{\rm M_{\odot}~yr^{-1}}$ in
the case of A1835. In a typical BCG it is one to ten per cent of the
mass cooling rate inferred from the hotter X-ray emitting gas assuming
it cools completely. UV-excess BCGs probably have the highest rate of
star formation of early-type galaxies at the present epoch. The
brightest galaxy in non-X-ray-peaked clusters do not have this
emission.

The excitation of the emission-line nebulosity, which often has a
filamentary structure (e.g. the Perseus cluster: \citett[conselice00]:
the Centaurus cluster: \citett[crawford05]; the Virgo cluster:
\citett[sparks04]) is likely related to the star formation but in
detail there are problems. Hotter emission is required than expected
from the stars observed and there are no stars obvious within the
outer filaments. The gas, even in the outer filaments, contains dust
and molecular hydrogen \cite{hatch05a,jaffe05}. Simple models for the
molecular gas imply higher pressures than for the surrounding gas and
very short radiative cooling times \cite{jaffe01,wilman02}. The
filaments may have been pulled out from a central reservoir of cold
and warm gas by the action of radio bubbles
\cite{fabian03b,hatch05b}. Where the central gas comes from is
not clear although radiative cooling is likely. A residual cooling
flow is therefore taking place in many clusters. 

Gas at intermediate temperatures of about $10^{5.5}~{\rm K}$ is seen through
OVI emission with FUSE. \citeextra[Oegerle et al.;oegerle01] found
such emission from A2597 and recently it has been found in A426 and
A1795 \cite{bregman05}. The inferred cooling rates in the 30 arcsec
FUSE aperture are $30-50~{\rm M_{\odot}~yr^{-1}}$.

\clearpage

\section{Heating}

Some heating is always expected in the central regions of clusters.
Examples are supernovae \cite{silk86,domainko04}, an active central
nucleus \cite{bailey82,tucker83,pedlar90,tabor93,binney95} and many
more recent papers cited in Sec 7.2--7.5), conduction
\cite{takahara79,binney81,stewart84,friaca86,bertschinger86,rosner89}
and many more recent papers cited in Sec 7.1).

A problem with heating the gas is that the cooling rate is
proportional to the density squared whereas most heating processes are
proportional to volume. This tends to make the gas unstable and means
generally that the cooler denser gas will carry on cooling while
hotter surrounding gas heats up. The gas appears to cool by about a
factor of three and then stop cooling. A mechanism to do that is not
obvious, since the gas does not appear to be piling up at the lower
temperature. Indeed it seems that the gas temperature profile is
"frozen" and has been so for some Gyrs \cite{bauer05}. 

The profile of $t_{\rm cool}$ is similar in many clusters with a
common central minimum value for $t_{\rm cool}$ of about 200~Myr
(Fig.~6). This
strongly suggests that heating is continuous, at least on timescales
of $10^8~{\rm yr}$ or more and is spatially distributed. Moreover, no shock
waves are apparent in these regions so any mechanical energy injection
must be subsonic.

\citeextra[Brighenti \& Mathews;brighenti03] (see also \citett[ruszkowski02])
provide several 1D simulations of a cluster core in which various
levels of heat are injected at various radii. The results which best
fit the temperature profile of a real cluster are those without
heating. This demonstrates that a heating model is not entirely
straight-forward. 

Some authors use a combination of processes. Although several are
operating in any cluster, it is unlikely that they have similar
weight, especially when it is recalled that the cooling flow problem
exists in objects ranging from elliptical galaxies, through groups to
the most massive clusters. It applies to a range of about 100,000 in
X-ray luminosity and over 15 in temperature. We are therefore seeking
a wide ranging, quasi-continuous, gentle, distributed heating process

We now consider conduction and AGN heating in detail before discussing
other models.

\subsection{Heat conduction}

The outer cluster atmosphere beyond the cooling radius represents a
vast reservoir of thermal energy. Conduction can transfer heat into
the cooler central regions.  The rate of conduction is however highly
uncertain due to the presence of magnetic fields which are probably
tangled.

Most workers have adopted Spitzer \cite{spitzer62} conductivity ($\kappa_{\rm S}\propto
T^{5/2}$), which is appropriate for an unmagnetized gas, and then
assumed a suppression factor $f$. \citeextra[Narayan \& Medvedev;narayan01] suggest
that $f$ is about 1/5 since field lines wander away from each other
exponentially when tangled (see \citett[tribble89] for another discussion). 

\begin{figure}[ht]
\begin{center}
\includegraphics[width=0.9\columnwidth,angle=-90]{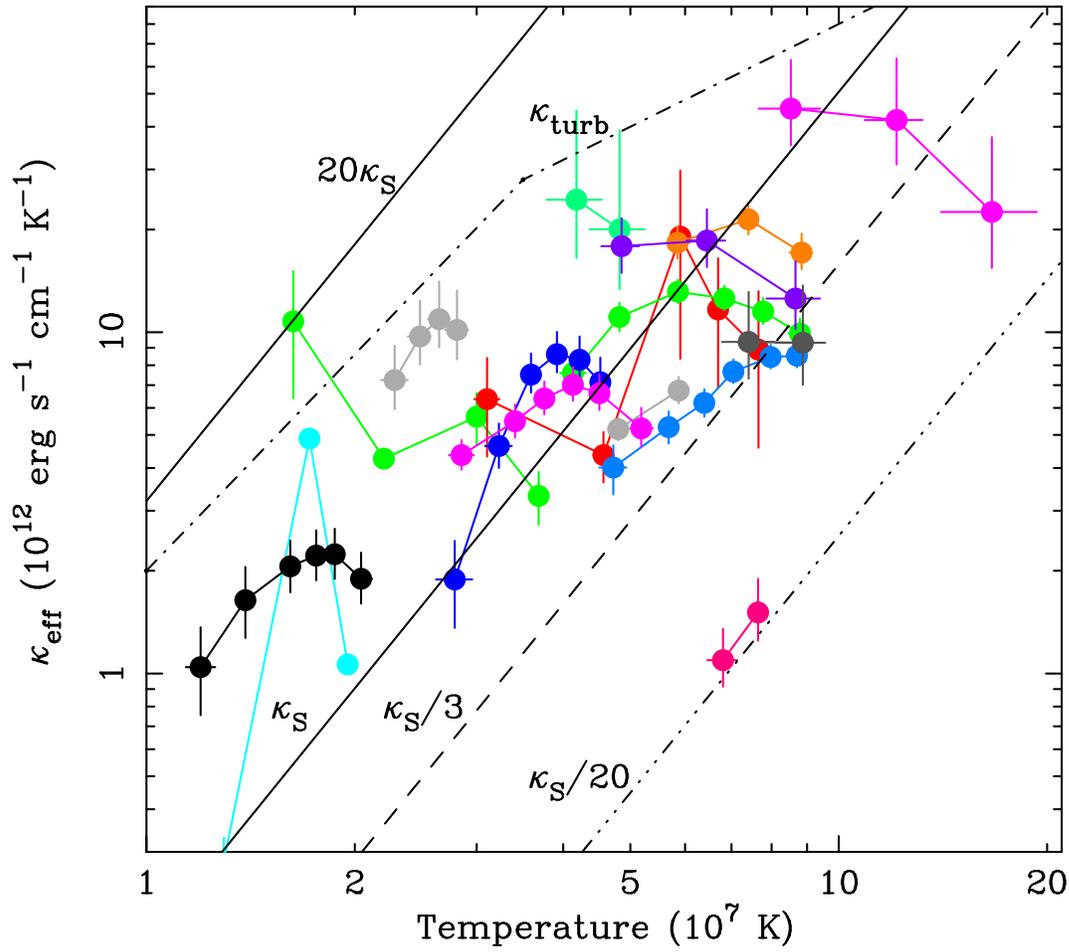}
\end{center}
\caption{The coefficient of electron thermal conduction that would be required
to balance X-ray cooling in the cores of a sample of clusters of galaxies.
These data are derived by measuring a mean temperature and density as a
function of radius and calculating the conduction coefficient to cancel
radiative cooling.  Each color is a different cluster and each point is a
measurement at various radii.  The lines are prediction for the magnetic field
free conduction coefficient of \citeextra[Spitzer;spitzer62].  It is unclear what to expect
for the effective conductivity in a cluster of galaxies with a complex
magnetic field topology (see text), but it would probably be below the Spitzer
value.  These results probably indicate the conduction is not strong enough to
cancel cooling by itself, but the value required curiously happens to be
within an order of magnitude of the canonical value.}
\end{figure}

Comparison with the data initially looked promising \cite{zakamska03,voigt02} then under further inspection ruled it
out \cite{kim03,voigt04}.  \citeextra[Voigt et al.;voigt02] (see also \citett[kaastra04] and \citett[voigt04]) used the density and temperature
profiles to deduce the heat transfer rate within the gas and thus the
effective conductivity $\kappa_{\rm eff}$ required. For several high
temperature clusters, $\kappa_{\rm eff}<\kappa_{\rm S}$ , but for
clusters where most of the gas is below 5~keV then $\kappa_{\rm
eff}>\kappa_{\rm S}$.  Then seems unlikely, however, but this may be possible \cite{cho04}.

\citeextra[Cho et al.;cho03] proposed that turbulent conduction might work,
whereby large subsonic eddies cause heat to be transferred in
radius. This does appear to operate fast enough to explain the data
\cite{kim03,voigt04} provided the gas is highly turbulent.
We shall return to the issue of turbulence later when discussing
heating by the central radio source (\S\ref{sec:radioheating}).

We note that conduction as a solution has a fine tuning problem
\cite{nulsen82,bregman88}. If the gas starts out
isothermal, then either the gas cools in which conduction cannot stop
it (conduction has a steep positive temperature dependence) or
conduction does work and there is no cool gas. How a profile
resembling a cooling flow would occur is difficult to understand.

\subsection{Heating by a central radio source\label{sec:radioheating}}

Most central cluster galaxies where $t_{\rm cool}$ in the surrounding gas 
is less than a few Gyr have a nucleus radio source \cite{burns90}. The
radio luminosity of these sources has a wide range from the powerful
FR\,II objects Cygnus~A and 3C295  to weaker FR\,I objects such as
M87. ROSAT HRI data clearly showed that the FR\,I lobes of 3C84, the
central radio source in the Perseus cluster, have displaced the ICM
and so are strongly interacting there \cite{boehringer93}. The
higher spatial resolution of Chandra has revealed further details of
that source \cite{fabian00,fabian03a,fabian05b} and enabled the discovery of
many more, similar, interactions. 
\begin{figure}[ht]
\begin{center}
\includegraphics[height=0.9\columnwidth,angle=-90]{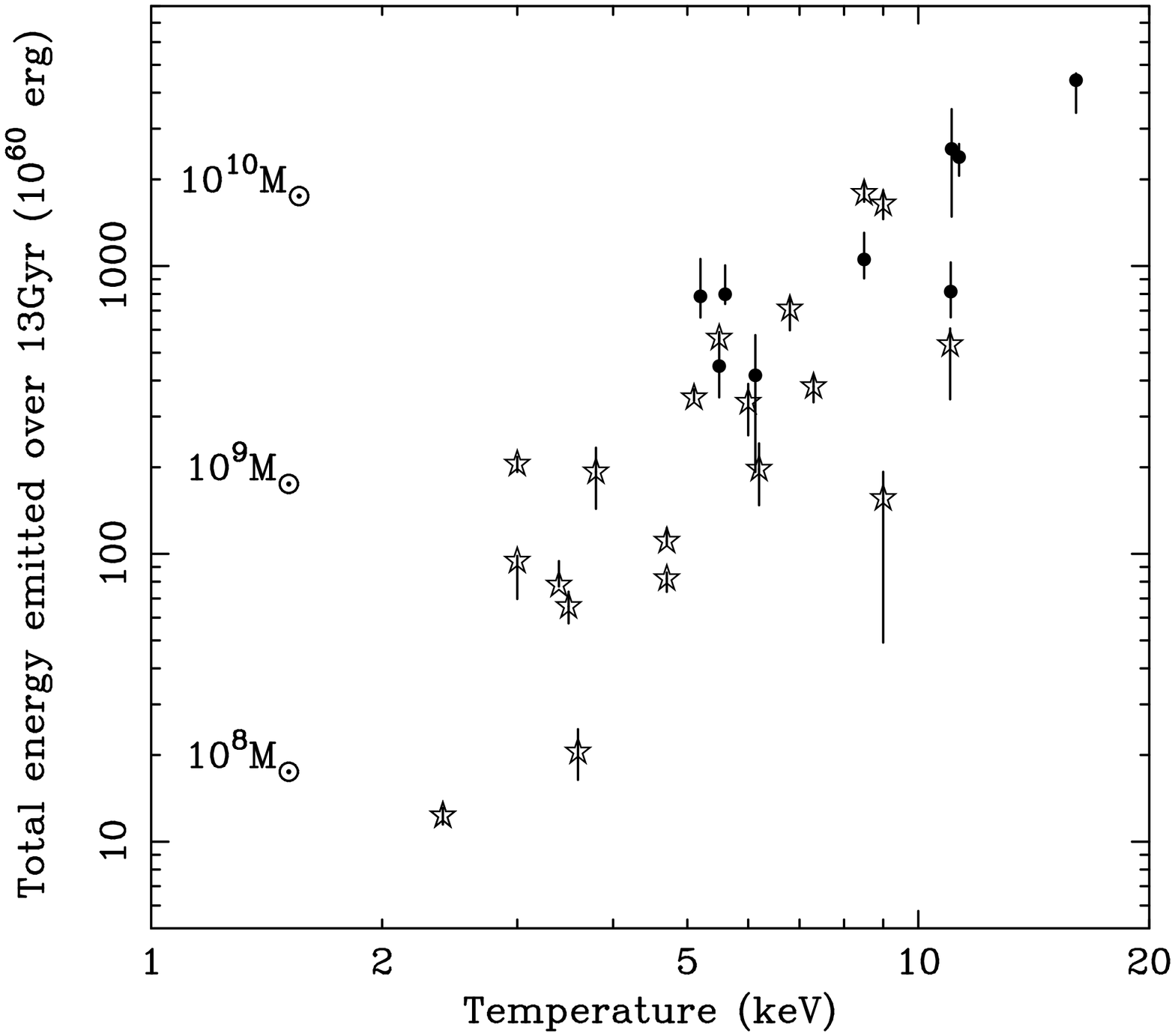}
\end{center}
\caption{The total heating power required to balance radiative cooling a
sample of cooling-flow clusters over a Hubble time.  The left hand side also
indicates the mass of the black hole required if ten percent of its rest mass
were applied directly into heating the ICM.  Figure adapted from 
\citeextra[Fabian et al.;fabian02].  }
\end{figure}

\begin{figure}[ht]
\begin{center}
\includegraphics[height=0.9\columnwidth,angle=-90]{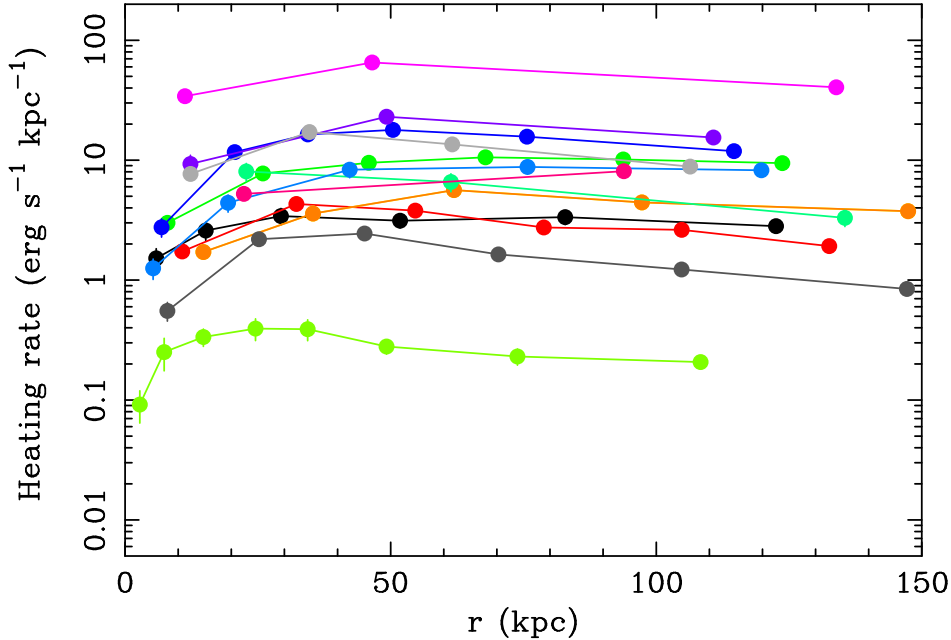}
\end{center}
\caption{The instantaneous heating rate required as a function of radius for a
sample of massive cooling-flow clusters.  These results indicate that the
heating has to be spatially distributed over a substantial physical volume.
Figure adapted from \citeextra[Voigt et al.;voigt04].}
\end{figure}

\subsubsection{Overview of radio bubbles}

Holes in the X-ray surface brightness coincident with radio lobes are
commonly seen and generally referred to as radio bubbles. They are
presumed to be relatively empty of normal thermal gas (see limits by
Schmidt et al \citett[schmidt01]) and mostly filled with a plasma
consisting of relativistic electrons, protons and magnetic field.
Striking examples apart from in the Perseus cluster are seen in
Hydra~A \cite{mcnamara00}; A2052 \cite{blanton01}; Centaurus
\cite{fabian00}; M87 \cite{young02,forman03}; A2597
\cite{mcnamara01,clarke05} and many more \cite{birzan04,dunn04,dunn05}.

Sometimes outer bubbles are seen (the NW one in the Perseus cluster
was first seen in images from the Einstein Observatory,
\cite{fabian81,branduardi81} which are considered to be ghost lobes.
Synchrotron and other losses have depleted the population of
radio-emitting electrons to the extent that they are not detectable in
the radio (particularly not at high frequencies) yet the other
electrons, protons and magnetic field in the cavities still exclude
the X-ray emitting thermal gas. Weak, low-frequency radio emission
pointing at the outer bubbles in the Perseus cluster supports this
hypothesis \cite{fabian02}. 

The bubbles discussed so far have radii from about 1 to 15~kpc. 
Recently, giant bubbles of radius 100~kpc have been found 100-200~kpc
from the centers of the Hydra-A nucleus \cite{nulsen04} and in
MS0735.6+7421 \cite{mcnamara05}, and some bubbles with little energy
(e.g. \citeextra[Blanton et al.;blanton04]). The minimum energy of a bubble can be
estimated from its volume $V$ and surrounding pressure $P$ as $E_{\rm
bubble}=PV$.  If the interior pressure is entirely relativistic in gas
with a ratio of specific heats of 4/3 then $E_{\rm bubble}=4PV$. The
energies are significant for typical bubbles \cite{birzan04} and
exceed $10^{61}~{\rm erg}$ for the giant bubbles.

\subsubsection{The simple theory of bubbles}

Bubbles from radio sources were first predicted by Gull \& Northover
\cite{gull73} and studied analytically before the Chandra/XMM-Newton
era by \citeextra[Churazov et al.;churazov00,churazov01] (see also
\citett[soker02]) and simulated by \citeextra[Heinz et al.;heinz98]. They
are inflated by jets from the nucleus and typically form in pairs
either side of the nucleus and seemingly attached to the nucleus. If a
jet of power $L$ steadily inflates a bubble of radius $R$ then,
after an initial supersonic phase and assuming radiative losses are
negligible, the bubble expansion with time $t$ follows from the
conservation of energy as $Lt = P V \propto P R^3$. Thus
$R\propto t^{1/3}$ and the expansion speed of the bubble $v_{\rm
bubble}\propto t^{-2/3}$. The bubble is of much lower density than its
surroundings so is buoyant and rises at a velocity close to the local
gravitational free-fall value $v_{\rm grav}$. When $v_{\rm bubble}$
drops below $v_{\rm grav}$ the bubble detaches from the jet and
rises. A new bubble then forms if the jet remains powered.

A simple estimate of the power injected into the bubbles can be
obtained by dividing $E_{\rm bubble}$ by an estimate of the age of the
bubble, which can be obtained from the above scaling (say, $R/v_{\rm
grav}$) \cite{birzan04,dunn04,dunn05}. These are typically several orders of
magnitude higher than the inferred radio luminosity of the jets and
nucleus showing that the radio luminosity is a poor guide to the power
of a jet. Such radio jets and lobes can have a radiative efficiency
(ratio of observed synchrotron luminosity to mechanical power) as low
as $10^{-3}$ or even $10^{-4}$. This, together with time variability
of the jets, indicates that the very poor correlation between the
radio power of the central source in a cluster and the heating power
\cite{voigt04,kaastra04} does not necessarily imply that the {\it mean} jet
power does not correlate well.

The power of the jets is mostly sufficient to offset radiative cooling
in the cluster core provided it is steady. \citeextra[Birzan et al.;birzan04] did
however note that it is insufficent in some objects. This could
indicate that radio source heating is not a general process or that it
can vary by factors of a few on the bubbling timescale, which ranges
between $\sim5$~ Myr and  $\sim 50$~Myr. This timescale is still
shorter than the inner radiative cooling timescale which is typically 
100--300~Myr. 

A further problem they raise is that bubbles were only found in 10 per
cent of the clusters they examined, so they either do not occur in
many clusters or if they do then they do not occur often. This result
could however be a selection effect from the examination of a sample
of clusters mixed in type (i.e. whether the cooling flow problem
applies to that cluster or not) and signal-to-noise. A simple
inspection by ACF of Chandra images of the 30 per cent of the 55
brightest clusters in the Sky listed by \citeextra[Edge et al.;edge92]
which have central $t_{\rm cool}<5$~Gyr shows that all but one have
clear bubbles, so we suspect that bubbles are indeed common enough to
be a viable ingredient in the solution to the cooling flow problem. A
more detailed analysis of the issue by \citeextra[Dunn et al.;dunn05] shows
that at least 75 per cent of the cluster cores needing heating have
bubbles. This means that the duty cycle of bubbles is such that they
are (detectably) present for 75 per cent of the time. 

Finally, before examining numerical simulations and then the heating
mechanism in detail, we note that the total energy required to stave
of cooling for several Gyr in a luminous cluster implies a large
central black hole mass (\citett[fabian02]; assuming a mass-to-energy
conversion effiency of 10 per cent). \citeextra[Fujita \& Reiprich;fujita04]
compared black hole masses predicted by the velocity dispersion relation \cite{tremaine02} and
found that they often fall short of what is required. However for the
very few systems where the mass has been measured (e.g. M87) there is
no problem. Perhaps the conclusion here would be that massive black
holes in cluster cores lie above the $M_{\rm BH} - \sigma$
relation for lower mass galaxies.

\subsubsection{Numerical simulations of radio bubbles}

Many groups have now carried out numerical hydrodynamical simulations
of the behaviour of bubbles in the
ICM \cite{bruggen01,quilis01,bruggen02,bruggen03,basson03,omma04a,omma04b,ruszkowski02,ruszkowski04a,ruszkowski04b,robinson04,vecchia04,reynolds05}.

Most have been 2D or 3D and use the FLASH or ZEUS codes (see
\citett[gardini04]) for a brief discussion and comparison of many of the
simulations). Several assume that the surrounding gas is isothermal.
Many produce bubbles which are very unstable (to Rayleigh-Taylor and
Kelvin Helmholtz instabilities depending on the motion of the bubble
at the start of a run) and collapse once they have risen more than
their own height, which is most unlike the observed bubbles. Indeed
most simulated bubbles do not look like the observed ones.

Several of the early simulations relied on the dragging out of cooler
gas from the smallest cluster radii by the bubbles as the explanation
of why so little cool gas is seen in real clusters. This explanation
however only works over the lifetime of the cluster (say 5~Gyr) if the
cooler gas is dragged out beyond the cooling radius (i.e well beyond
100~kpc), otherwise it either falls back in or mixes in and reduces
the cooling time of gas just beyond the center. A catastrophic cooling
flow is only being postponed for a while. Although this explanation
can work temporarily it cannot provide a comprehensive
solution.

When heating is estimated in simulations it is often unclear what it
is due to. Some work is done by the buoyant bubbles rising and the
sinking of cooler gas. This is the heating mechanism of the
'effervescent heating' approach of \citeextra[Begelman;begelman01]
(see also \citett[ruszkowski04a]).  $P d V$ work is done as the
bubbles are made in the first place. That energy presumably propagates
as a sound wave and need not be dissipated locally unless shocks are
involved or the ICM is viscous.

Heating by a mixture of weak shocks (at small radii) and viscous
damping of sound waves (at larger radii) has been proposed by
\citeextra[Fabian et al.;fabian03b] on the basis of the ripples seen
in deep Chandra images of the Perseus cluster (see also Forman et al
2004 for a discussion of ripples in the Virgo cluster).  Viscosity has
been included in simulations by \citeextra[Ruszkowski et
al.;ruszkowski04a,ruszkowski04b] and by Reynolds et al. \cite{reynolds05}
and appears to dissipate 20--30 per cent of the $P d V$ energy from
the bubbles. Transport of the energy by sound waves and dissipating it
over a lengthscale of 50--100~kpc provides a fairly gentle and
distributed source of heat. The bubbles in a viscous medium also
better resemble the observed ones \cite{reynolds05}.

\citeextra[Fujita \& Suzuki;fujita05] and \citeextra[Mathews et al.;mathews05] argue that strong
sound waves will shock and heat the innermost region and not deposit
energy further out where it is needed. However a very deep image of
the Perseus cluster \cite{fabian05b} shows that the shock (Fig.~15)
is isothermal. This may indicate that thermal conduction operates at
least within the inner parts of the cool region to share the energy
released by the bubbling process.

\subsection{Is the ICM turbulent?}

Several workers \cite{cho03,inogamov03,chandran04,fujita04} suggest
that the ICM is turbulent.  Such motions can transport heat and the
dissipation of the turbulence is itself a source of heat.
\citeextra[Schuecker et al.;schuecker04] have measured brightness and
temperature fluctuations in the Coma cluster (which is not a cooling
flow cluster).  \citeextra[Vogt \& Ensslin;vogt04] have determined a
limited turbulent-like spectrum in the Faraday Rotation Measures of
the Hydra cluster core.  \citeextra[Ensslin \& Vogt;ensslin05] have
extended their work to predict velocities and length-scales for the
turbulence in a number of cool core clusters.  They argue that a
small-scale turbulent dynamo is maintaining magnetic fields and that
turbulent dissipation can balance radiative cooling.

Contrary indications have been identified (Fabian et al 2003b) in the
highly extended, and often linear, optical filaments in the Perseus
\cite{conselice00} and other clusters (e.g A1795, \citett[cowie83];
Centaurus, \citett[crawford05]). Such filaments have coherent velocity
fields and small velocity
widths \cite{hu85,crawford05,hatch05b} so must be relatively old (perhaps
100~Myr) and in thermal pressure equilibrium with the surrounding ICM.
The optical surface brightness of such filaments translates to a
thickness of the emission regions of less than a pc. If the ICM is
turbulent and pushing on a filament, it will respond after interacting
with its own column density which means a kpc or so of ICM, given the
thousandfold higher density in the filament. This is only a few arcsec
in nearby clusters so the apparent linear nature of the filaments
would soon be destroyed by a fully turbulent ICM. They may however
reflect streamlines and could have been drawn out by the motion of
bubbles.  Note the similarity of the horseshoe-shaped filament in the
Perseus cluster to that of streamlines behind rising water bubbles; see
Fig. 184 in \citett[vandyke82] and discussion by \citeextra[Fabian et
al.;fabian03b]

The inner ICM in clusters with significant linear filaments may
therefore have subsonic random flows, particularly radial behind
bubbles. Whether this is turbulence depends on how turbulence is
defined. It does not resemble fully-developed hydrodynamical
turbulence (see Fig. 186 in \citett[vandyke82]) but is not dissimilar from
flows in some magnetohydrodynamical simulations (e.g.  \citett[schekochihin04]). 

What is unclear is the driving mechanism for turbulence in a cluster
core. A subcluster merger may well have made the Coma cluster ICM
turbulent, but such mergers will not have strongly affected the dense core
cluster under discussion here (but see \citett[fujita04]).  Driving
turbulence into the dense core will require considerable energy.   Rising radio bubbles are a possible
driver but whether their motion leads to a turbulent cascade of energy
to much smaller scales must be determined. The viscosity of the ICM
may play a key role in this issue.

The abundance gradients seen at the centers of many cool core clusters
provides a further constraint on turbulence and diffusion.
\citeextra[Rebusco et al.;rebusco05] have modelled the production and
maintenance of the gradient in the Perseus cluster and find a
diffusion coefficient $D \sim v \ell \approx 2\times
10^{29}$~cm$^2$~s$^{-1}$. \citeextra[Graham et
al.;graham05] have carried out a similar analysis on the more
abundance-peaked Centaurus cluster to obtain a value of $D \sim 5$
times smaller.

\subsection{Multiphase flows}

Large scale global flows have been considered that transfer significant
thermal energy from one region of the cluster to another.
\citeextra[Mathews et al.;mathews03a,mathews04] have discussed flows
which move in both radial directions. To be long lived, both mass and
energy must be supplied to the inflowing gas over a large volume. The
energy is assumed to be derived from bubbles.

\citeextra[Fabian;fabian03a] considered a multiphase flow in which outer denser blobs
fall inward and mix with inner, cooler blobs, thereby tapping
the extended gravitational potential. Unless there is some outflow or cooling,
however, mass builds up near the cluster center.

\subsection{Role of Magnetic Fields and Cosmic Rays}

A cooling flow will amplify tangled magnetic fields in the
intracluster gas \cite{soker90}. Magnetic fields in the
general ICM have been found through radio observations of Faraday
Rotation \cite{kim01,clarke04}. Such observations also show
it rising in cool core clusters where the highest values have been
found \cite{carilli02}. The results indicate a roughly
cellular structure for the magnetic field of a few kpc (e.g. the
Centaurus cluster, \citett[taylor02]). Such a structure may occur from
a spectrum of magnetic fields perhaps due to turbulence
\cite{ensslin05}. Near the center the magnetic pressure can be 10 per cent
of the thermal pressure, perhaps more. An important effect of the
magnetic field is that it can dramatically reduce the microphysical
transport processes in the ICM.

Cosmic rays are a likely additional component in cluster
cores. Relativistic electrons produce the minihalo in the X-ray peaked
region of the Perseus cluster (see \citett[gitti04] for a theoretical
discussion) and may produce a hard X-ray component by inverse Compton
scattering \cite{sanders05}. The effects of cosmic rays could be
important in mediating the flow of mechanical energy in the core.

\citeextra[Cen;cen05] notes that a cosmic ray phase can make heated gas thermally
stable. A floor temperature of 0.3 times the local ambient temperature
is derived if the cosmic ray pressure is 1/3 of the ambient value.
This situation is different from the similar factor seen in
observations, which applies across the whole cool region, not just
locally. Note that gas is not observed to "pile up" at any particular
temperature.

If cosmic rays are spread through the core in many small bubbles, then
they would be very effective in dissipating the energy in sound waves
(Heinz \& Churazov 2005). Such bubbles must be small not to be
detectable in deep images (eg. Fabian et al 2005b).

\subsection{Feedback}

If AGN heating balances cooling on timescales of $\sim 10^8$ to a few
$10^9~{\rm yr}$ then some feedback is needed to prevent either a cooling
catastrophe (none has yet been found) or an event heating all the
central gas and blowing it out. Bondi accretion onto a central black
hole provides a link between the two regions but the radius range from
the Bondi radius of say 50~pc to the ICM at 50~kpc is a factor of one
thousand. The volume occupied by the black hole and its accretion
radius which is required to provide the heating is one part in a
billion!

Such numbers emphasize the problem faced by the necessary feedback. 
Some attempts to model feedback are discussed by \citeextra[Ciotti \&
Ostriker;ciotti01]; \citeextra[Kaiser \& Binney;kaiser03];
\citeextra[Nulsen et al. 2004;nulsen04], \citeextra[Soker \&
Pizzolato;soker05], \citeextra[Omma \& Binney;omma04a];
\citeextra[Binney;binney04]; \citeextra[Hoeft \& Br\"uggen;hoeft04];
\cite{churazov05};\cite{pizzolato05}.
The common occurrence of star formation in the central galaxy and also
of giant bubbles shows that any feedback need not be perfect on
timescales less than $10^8~{\rm yr}$. \citeextra[Donahue et al.;donahue05] study two
clusters which show little central temperature drop or central radio
source yet have central cooling times of $\sim 1$~Gyr. Such clusters,
which are in the minority, may be recovering from giant outbursts.

\subsection{Other heat sources and mechanisms}

Cosmic ray heating has been invoked in several models
\cite{colafrancesco04,totani04}. Generally to be effective, the
pressure in cosmic rays would have to exceed that of the thermal gas
\cite{loewenstein91}. Excess ionization by suprathermal electrons
could change the spectral appearance of gas in a cooling flow. This
has been studied by \citeextra[Oh;oh04], who finds that such electrons provide
much more heating than ionization, so reverting to the problems raised
above by \citeextra[Lowenstein et al.;loewenstein91].

Heating by galaxy motions have been revived by \citeextra[El Zant et
al.;elzant04] (see also \citett[faltenbacher05]) following early work
by \citeextra[Miller;miller86]. It may be a viable distributed source
of heat if the mass-to-light ration of the galaxies passing through
the inner regions exceeds 10 (but see \citett[kim05]).

Dark matter interactions have been discussed by several authors
\cite{qin01,totani04,chuzhoy04}.  Similarities with the solar corona
have been discussed by \citeextra[Makishima;makishima01] and \citeextra[Kaastra et
al.;kaastra04].  \citeextra[Br\"uggen \& Ruszkowski;bruggen05] discuss that viscous heating, if it occurs in real MHD plasmas, could
provide significant heating during structure formation.

One last possibility is that sedimentation acts \cite{fabian77}.
Helium then accumulates at the cluster center \cite{gilfanov84}
possibly explaining the puzzling low metallicity found at the center
of some nearby clusters observed with Chandra. The main low temperature
cooling gas is then helium rich which can produce much weaker emission
lines. Stars formed from helium rich gas will have short lifetimes so
avoiding some constraints on the cold mass sink
\cite{fabian03c,ettori05}. Tangled magnetic fields will of course
significantly reduce the prospect of any sedimentation.

\clearpage

\section{Discussions}

The simplest explanation for the common appearance of cold core, X-ray
peaked clusters is that, when averaged over tens of Myr, the radiative
cooling is balanced in part by distributed heating. Thermal conduction
as a means of distributing heat from outer gas is ruled out for low
and intermediate temperature clusters. It may however have a role in
spreading the energy in the central parts.  A plausible mechanism is
the dissipation of energy propagating through the ICM from a central
radio source. Such a process stems massive cooling onto the BCG which
would otherwise gain a total stellar mass $\gg10^{12}~{\rm M_{\odot}}$. The
process is therefore a vital ingredient in stopping the growth of the
most massive galaxies \cite{fabian02b,benson03,binney04}. NOte that
most semi-analytic models for galaxy formation (e.g. Kauffmann et al
1999) already needed to suppress cooling in massive haloes in order to
match observation.

Difficulties and doubts remain with regard to the issues of the energy
dissipation and distribution processes which are tied in with the
transport processes in the gas. Similarly, it is not clear how the
feedback manages to produce such similar cooling time profiles in
systems where temperatures and thus masses differ by over an order of
magnitude?  There still remains the possibility that some process not
yet foreseen, or at least not well studied will eventually prove more
important than the effect of the central radio source, or will at
least be important in mediating its effect. The central radio source
is so common and so energetic however that it must at least be part of
the solution. Similarly, the motions of galaxies or interacting dark matter,
if it exists, could be important in heating cluster cores along with the AGN.
Given the wide range of objects in which a balance is
required, we suspect that a single mechanism is dominant, rather than
several.

The need for a heating-cooling balance is in a time-averaged sense,
over intervals of about $10^8~{\rm yr}$. In most cases the heating has not
been so energetic as to drive gas out of the inner regions nor so weak
as to allow much cooling at very high rates. Examples of objects at
the extremes are Hydra~A and Cygnus~A where heating is high (but is
dumped mostly at large radii outside the cool region) and A2597 and
RXC1504.1-0248 where cooling appears to be high (in the latter object
over 70 per cent of the total X-ray luminosity emerges from the cool
core; \citett[boehringer05]). 

In most objects residual cooling at a rate of about 10 per cent of the
simple unheated cooling rates appears to occur. It could be larger if
non-radiative cooling, due to mixing say, is occurring. Stars form
from the cooled gas giving the excess blue light seen in the BCGs.
Mass loss from such stars can make the cooled gas dusty and radio
bubbles drag some of it back out to large radii.

The time evolution of the heating / cooling balance is little
understood. We suspect that the common temperature drop associated
with the short central cooling times {\em is} due to radiative cooling
and that heating only came into balance when the overall temperature
structure was in place. Perhaps the central galaxy and its central BH
grew until balance was achieved, and growth required cooling.
Comparison of samples of clusters at $\bar z=0.22$ with $\bar z=0.056$
\cite{bauer05} shows little surprisingly change in the
distribution of cooling times at 50 kpc. The results imply that any
balance was established well beyond $z\sim 0.3$.

Few massive cool core clusters are known at much higher redshifts than
that of RXJ1347 at $z=0.44$. This may however be a selection effect.
They will be absent from X-ray cluster samples if the central BH is a
bright X-ray source such as a quasar. If a central quasar outshines
the host cluster in X-rays then the object will generally be
classified as a quasar. H\,1821+643 is a good example of a bright
quasar in an X-ray peaked cluster at intermediate redshift ($z=0.297$,
\citett[fang02]).  Searches for clusters around powerful radio-loud
quasars and galaxies have found some examples at $z=0.5-1.1$
\cite{worrall01,crawford03,siemiginowska05} but no complete searches
have been done.

\clearpage

\section{Future work}

Further detailed deep studies with Chandra and XMM-Newton as well as future
studies with Constellation-X and Xeus are needed
in order to better understand the heating/cooling balance. There is
considerable potential for studies of the cool and warm gas and dust
around the BCG using HST, Spitzer and ground-based telescopes. Larger
cluster samples are needed, particularly at medium to high redshifts.
Determining the extent of a heating-cooling balance (or not) in
groups, elliptical galaxies, and galaxy bulges is also important as is
the evolution of any balance in all massive objects.

\clearpage

\section{Acknowledgments}
We thank Mateusz Ruszkowski, Mitch Begelman, Roger Blandford, and an anonymous
referee for a careful reading of this manuscript.  ACF thanks the Royal
Society for support.  JRP was supported by a grant from NASA for scientific
and calibration support of the RGS at Stanford.  This was also supported by
the U.S. Department of Energy under contract number DE-AC02-76SF00515.

\end{document}